\documentclass[final,5p,times]{elsarticle}

\usepackage{lineno,hyperref}
\usepackage{graphicx}
\usepackage{amssymb}
\usepackage{siunitx}
\DeclareSIUnit\clight{\text{\ensuremath{c}}}
\DeclareSIUnit[number-unit-product = ]\percent{\char`\%}
\usepackage{pdfpages}
\usepackage{subcaption}
\usepackage{float}
\usepackage{amsmath}
\usepackage{color}
\usepackage{ulem}
\usepackage{footnote}
\usepackage{textcomp}
\usepackage{xcolor}
\usepackage{xspace} 
\usepackage{booktabs}
\usepackage{multirow}
\usepackage{todonotes}
\usepackage{wasysym}
\usepackage{tabularx}
\usepackage{tablefootnote}
\usepackage{threeparttable}
\newcommand{\comment}[1]{}

\newcommand{\figref}[1]{Fig.~\ref{#1}}
\newcommand{\Figref}[1]{Figure~\ref{#1}}	
\newcommand{\secref}[1]{Section~\ref{#1}}

\newcommand{\tabref}[1]{Table~\ref{#1}}

\newcommand{\ArCOtwo}{Ar-CO$_2$ (90-10)\xspace}

\newcommand{\ibf}{ion backflow\xspace}

\newcommand{\sh}{single-hole\xspace}
\newcommand{\Sh}{Single-hole\xspace}

\newcommand{\alphasource}{($^{239}$Pu+$^{241}$Am+$^{244}$Cm) }

\journal{NIM A}
\title{New (TH)GEM coating materials characterised using spectroscopy methods} 

\begin{document}
\begin{frontmatter}


\author[a]{B.~Ulukutlu\corref{cor1}}
\ead{berkin.ulukutlu@tum.de}

\author[b,c]{P.~Gasik}

\author[a]{T.~Waldmann}

\author[a]{L.~Fabbietti}

\author[a]{T.~Klemenz}

\author[a]{L.~Lautner}

\author[d]{R.~de Oliveira}

\author[d]{S.~Williams}

\address[a]{Physik Department, Technische Universit\"{a}t M\"{u}nchen, Munich, Germany}
\address[b]{GSI Helmholtzzentrum f\"{u}r Schwerionenforschung GmbH (GSI), Darmstadt, Germany}
\address[c] {Facility for Antiproton and Ion Research in Europe GmbH (FAIR), Darmstadt, Germany}
\address[d]{European Organization for Nuclear Research (CERN), Geneva, Switzerland}

\cortext[cor1]{Corresponding author}

\begin{abstract}
In this work GEM and \sh Thick GEM structures, composed of different coating materials, are studied. The used foils incorporate conductive layers made of copper, aluminium, molybdenum, stainless steel, tungsten and tantalum. The main focus of the study is the determination of the material dependence of the formation of electrical discharges in GEM-based detectors. For this task, discharge probability measurements are conducted with several Thick GEM samples using a basic electronics readout chain. In addition to that, optical spectroscopy methods are employed to study the light emitted during discharges from the different foils. It is observed that the light spectra of GEMs include emission lines from the conductive layer material. This indicates the presence of the foil material in the discharge plasma after the initial spark. However, no lines associated with the coating material are observed while studying spark discharges induced in Thick GEMs. It is concluded that the conductive layer material does not play a substantial role in terms of stability against primary discharges. However, a strong material dependence is observed in the case of secondary discharge formation, pointing to molybdenum coating as the one providing increased stability.


\end{abstract}
\end{frontmatter}

\section{Introduction}
\label{sec:intro}

The requirements of the new generation experiments in particle physics are driving factors for the development of new detectors. Novel devices must handle very high particle rates and meet the requirements of large experiments, which in particular concerns the substantial increase in active detector area. Among the new innovative detector technologies, Micro-Pattern Gaseous Detectors (MPGD) have become widely used in high-rate physics experiments and are foreseen at future facilities. In particular, Gas Electron Multiplier (GEM)~\cite{sauli1997gem} and its coarse version Thick Gas Electron Multiplier (THGEM)~\cite{chechik2004THGEM} are well established types of MPGD with high position, time and energy resolutions. Moreover, they offer high rate capability and are used in a variety of modern physics experiments (see a recent review in~\cite{sauli2020book}).

The key parameter for long-term operation of such detectors is stability against electrical discharges (sparks). Experience shows that in real experimental environments there is a non-zero probability of a spark development. If the size of the avalanche in the amplification region reaches the critical charge limit, a streamer may develop, leading to a spark discharge. The critical charge necessary to trigger a spark in a GEM structure varies, depending on the gas mixture, between 10$^6$ and 10$^7\,e$ (see~\cite{bachmann2002discharge, gasik2017charge}, for example). In the high rate environment of Minimum Ionizing Particles (MIPs), reaching 1\,MHz\,cm$^{-2}$, the probability of exceeding these limits is very low, down to 10$^{-12}$ per incident hadron for a conventional triple GEM tracker~\cite{cardini2012operational}. However, the critical charge may be easily exceeded in the presence of highly ionizing particles such as alpha particles, low-energy protons or products from electromagnetic or hadronic showers. High charge densities liberated in the closest vicinity of the amplification structure may significantly alter the detector stability by triggering a spark discharge. 

\begin{figure}[ht]
\centering
\includegraphics[width=0.9\linewidth]{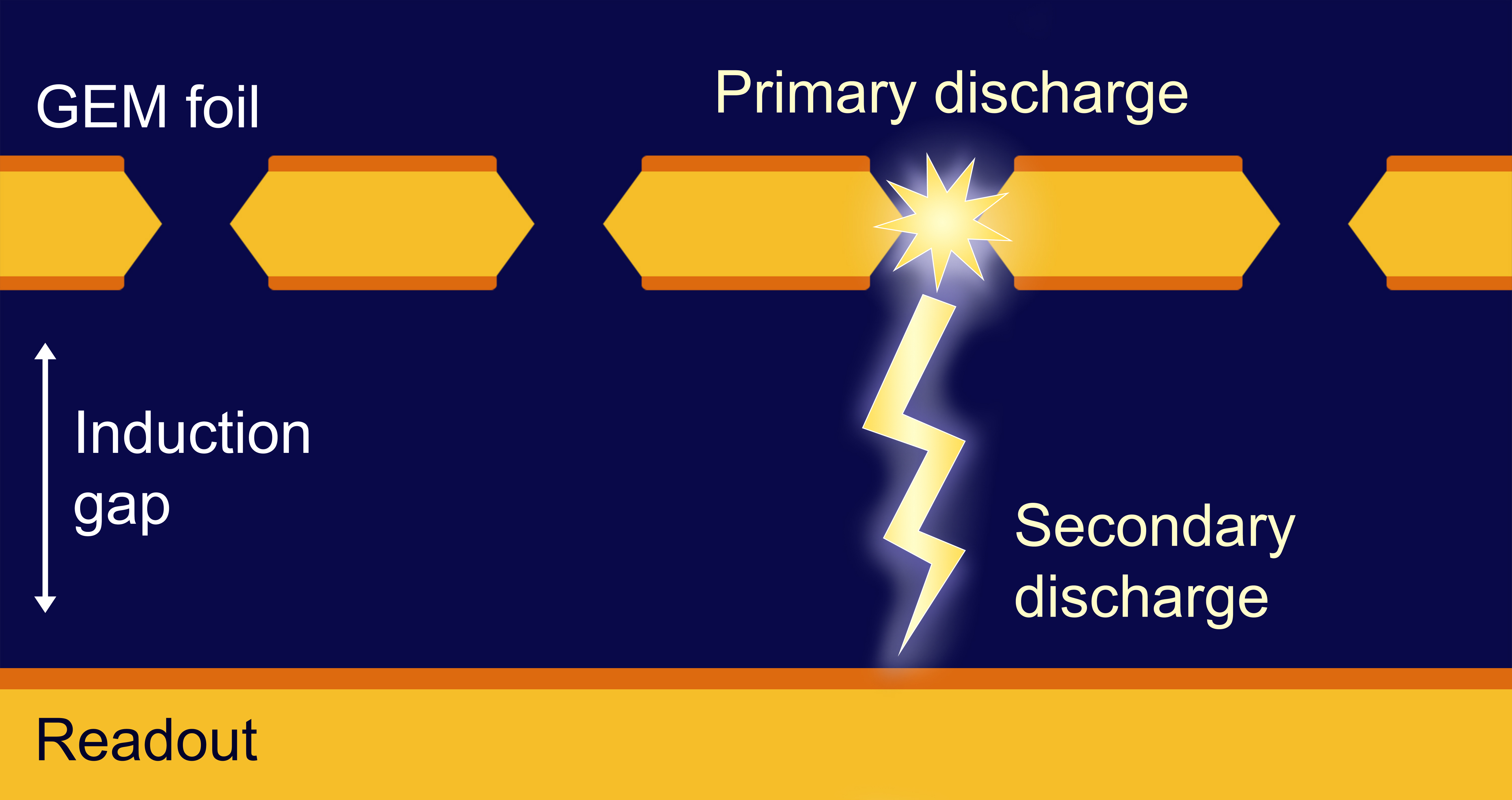}\\[2ex]
\includegraphics[width=0.9\linewidth]{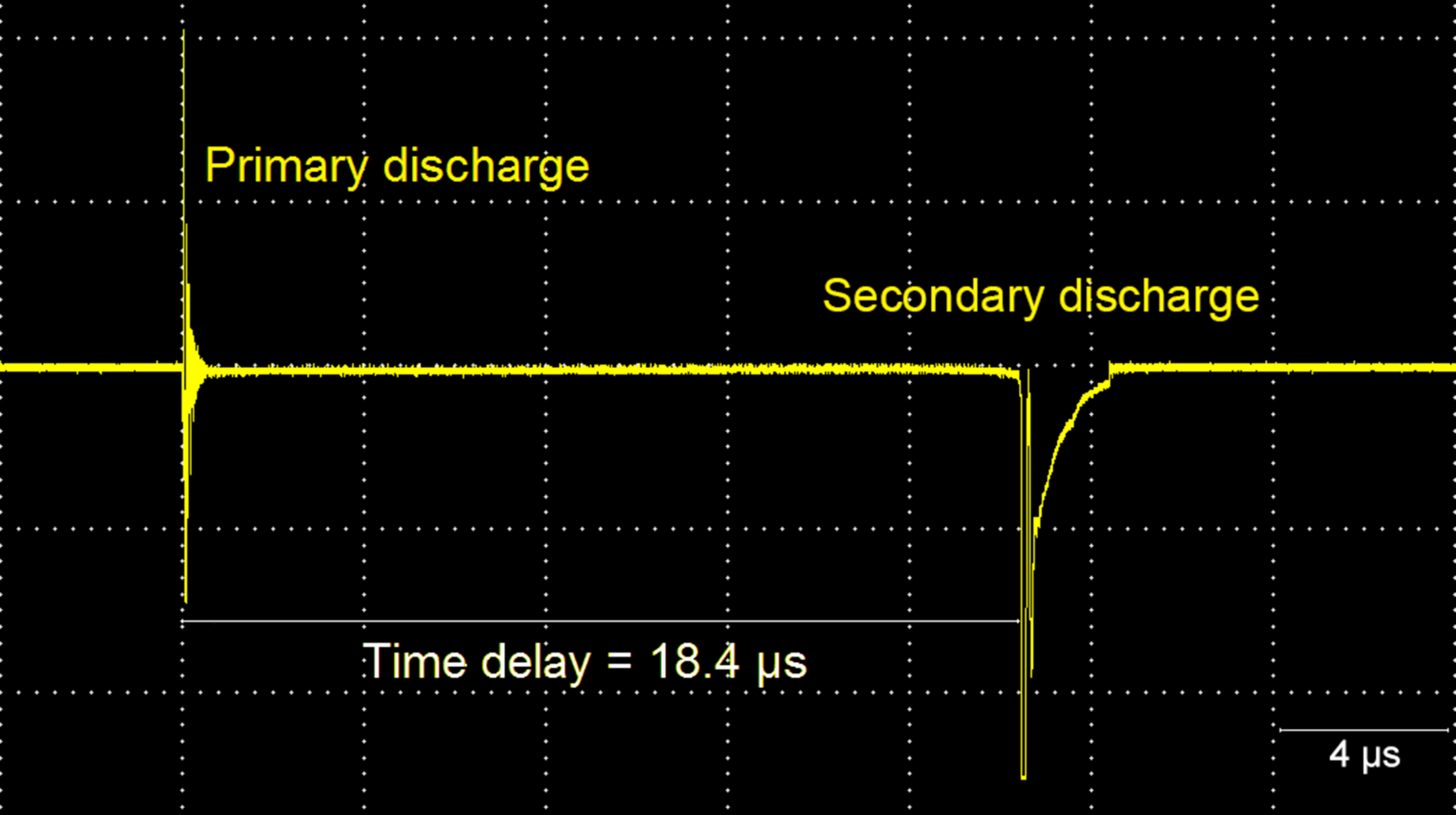}
\caption{(\textit{top}) Scheme of a GEM detector portraying a primary discharge that occurs inside a GEM hole, followed by a secondary discharge developed in the gap between the GEM and the readout electrode. (\textit{bottom}) Electrical signals induced by different types of discharges on the readout plane.}
\label{fig:primary_secondary}
\end{figure}

Another type of a potentially destructive event observed in GEM-like structures is called \textit{secondary discharge}\footnote{also dubbed in the literature as a propagated or a delayed discharge}. It can appear in the gaps between subsequent GEMs in a stack (transfer gaps) or between a GEM and a readout plane (induction gap), as schematically presented in \figref{fig:primary_secondary}. \cite{bachmann2002discharge, peskov2009research, deisting2019secondary, utrobicic2019studies}. 

Secondary discharges are triggered by sparks in the GEM holes (which in this context can be referred to as \textit{primary discharges}) and develop at the gap field values lower than the amplification field for a given gas mixture. The secondary discharge probability is usually a steep function of the electric field. A characteristic of this kind of discharges is the peculiar time lag between the primary and the secondary discharge reaching several tens of microseconds (see \figref{fig:primary_secondary}). The exact values of the onset field, discharge probability and the time lag depend, however, on the conditions at which the detector is operated, such as gas mixture, gap geometry, HV circuitry, and several others. The time lag, in addition, drops with increasing electric field in the gap \cite{deisting2019secondary}. An interpretation of the entire physical process leading to the creation of secondary discharges is thus complicated and still remains a subject of debate. In recent years, there were several approaches to unravel this process, which pointed to heating of the cathode upon a primary discharge as a possible mechanism of the secondary discharge creation \cite{deisting2019secondary, utrobicic2019studies}. Following the discussion in \cite{deisting2019secondary}, a primary discharge creates enough electrons to cause further gas ionisation and excitation in the gap below a GEM. It also heats GEM electrodes facilitating thermionic emission. The latter is sustained by the ion bombardment, causing massive secondary electron emission which may eventually transform into a spark.

Secondary discharges pose a threat to the integrity of the detector and its front-end electronics. The probability of secondary discharge occurrence must be reduced to a minimum by all means to assure reliable long-term operation. A number of mitigation strategies has been proposed \cite{deisting2019secondary, lautner2019high} and successfully implemented in the ALICE TPC \cite{TPCupgrade} and CMS muon spectrometer upgrades \cite{merlin2020cms}. It would still be beneficial to understand the entire process and, if possible, to eliminate the cause of these violent events completely.

In order to further investigate the mechanism of secondary discharge creation, we study (TH)GEM structures with electrodes produced out of diverse metals, characterised by different mechanical and thermal properties. We propose optical spectroscopy to examine the color spectra of the emitted light from primary and secondary discharges in order to investigate the content of plasma created in a discharge and its possible relation to the cathode heating process. In this work, we have conducted a comprehensive spectroscopy study using a number of GEM and THGEM foils operated under various gas mixtures. 
The measurements aim to investigate the effect of the used GEM foil materials on the operational stability of the detector against secondary discharges.
\section{GEMs and THGEMs with new coating materials}
\label{sec:GEMs}

\begin{figure}
     \centering
     \begin{subfigure}[b]{0.32\linewidth}
         \centering
         \includegraphics[width=\textwidth]{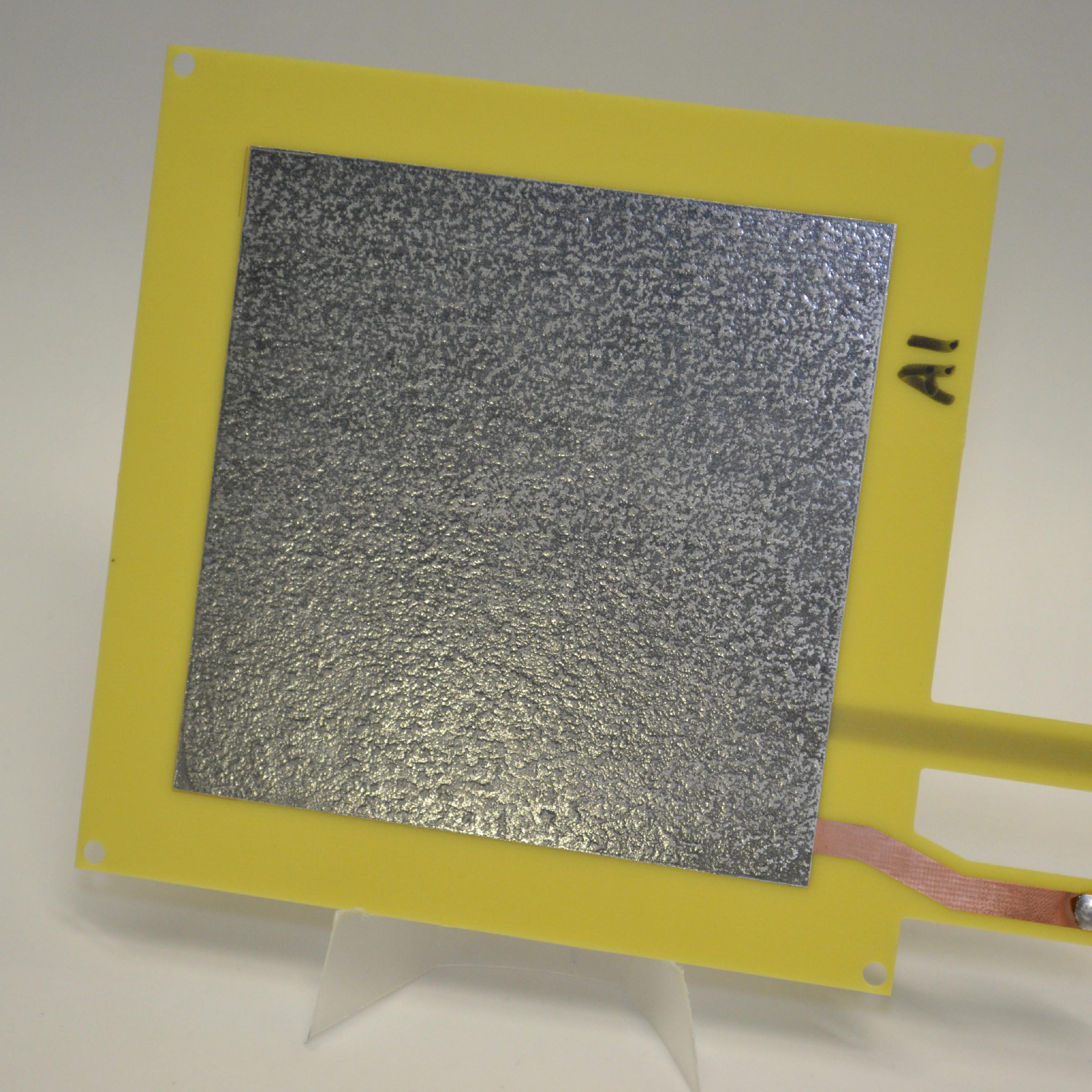}
         \caption{Aluminium}
         \label{fig:alu_pic}
     \end{subfigure}
     \hfill
     \begin{subfigure}[b]{0.32\linewidth}
         \centering
         \includegraphics[width=\textwidth]{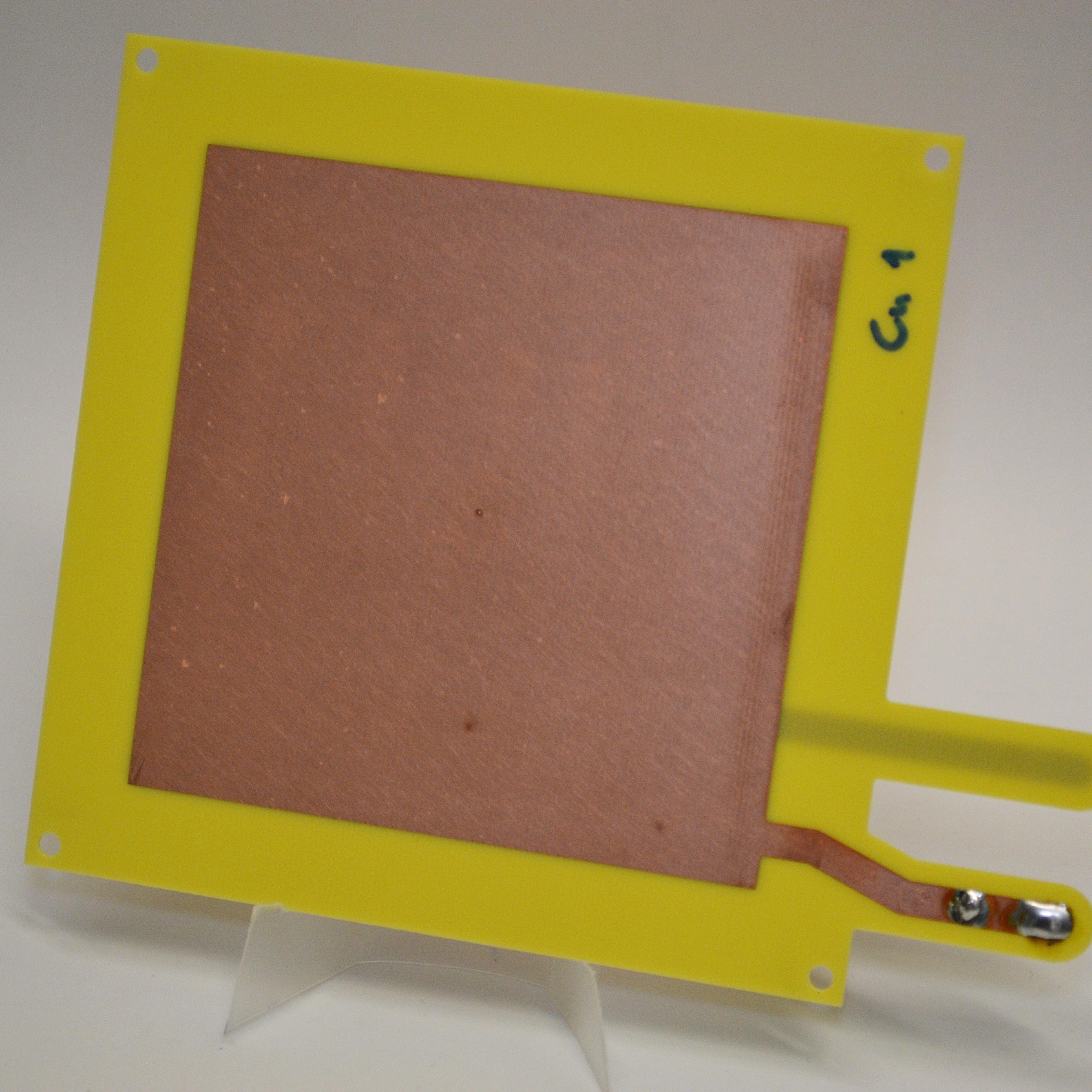}
         \caption{Copper}
         \label{fig:cu_pic}
     \end{subfigure}
     \hfill
     \begin{subfigure}[b]{0.32\linewidth}
         \centering
         \includegraphics[width=\textwidth]{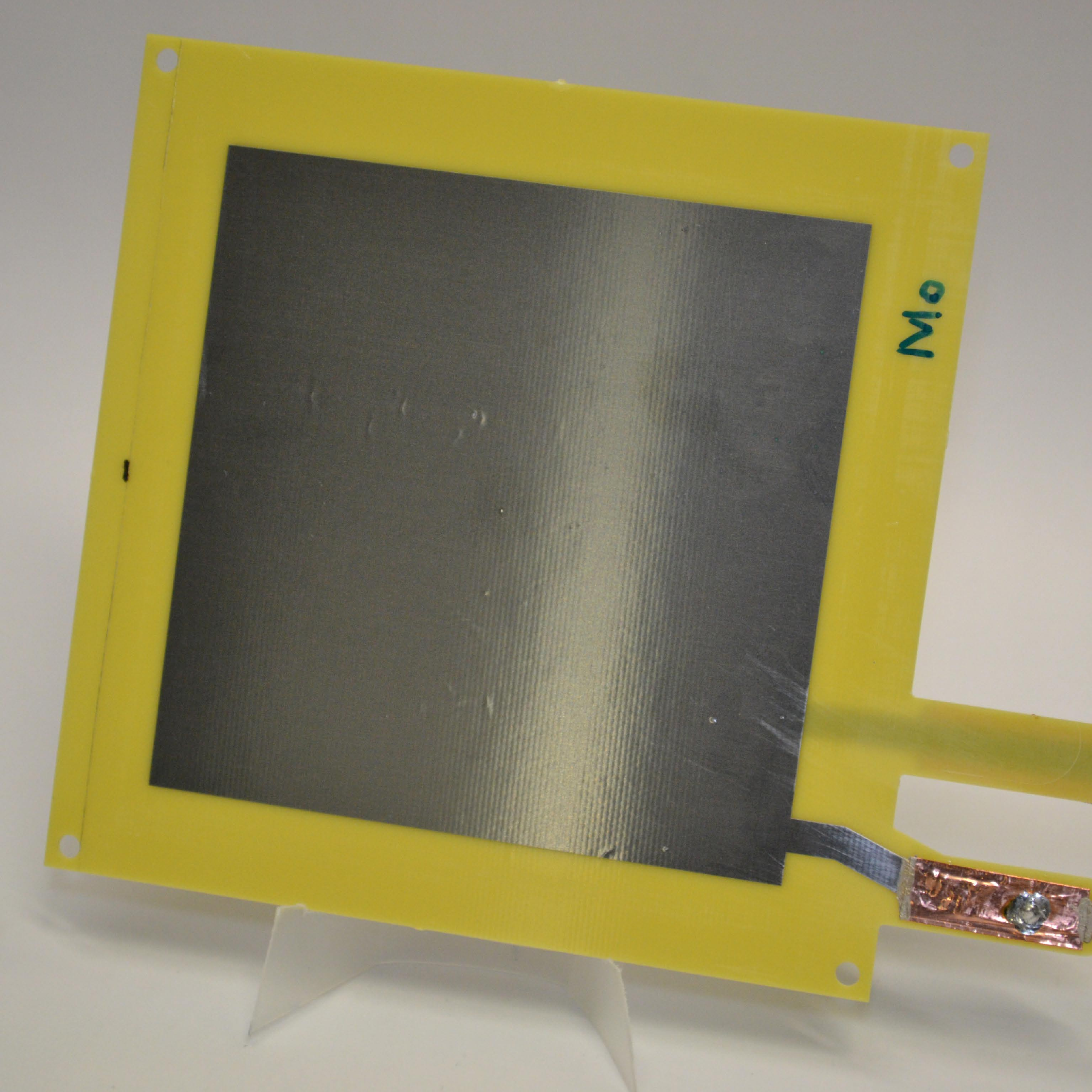}
         \caption{Molybdenum}
         \label{fig:mo_pic}
     \end{subfigure}
     \hfill
     \begin{subfigure}[b]{0.32\linewidth}
         \centering
         \includegraphics[width=\textwidth]{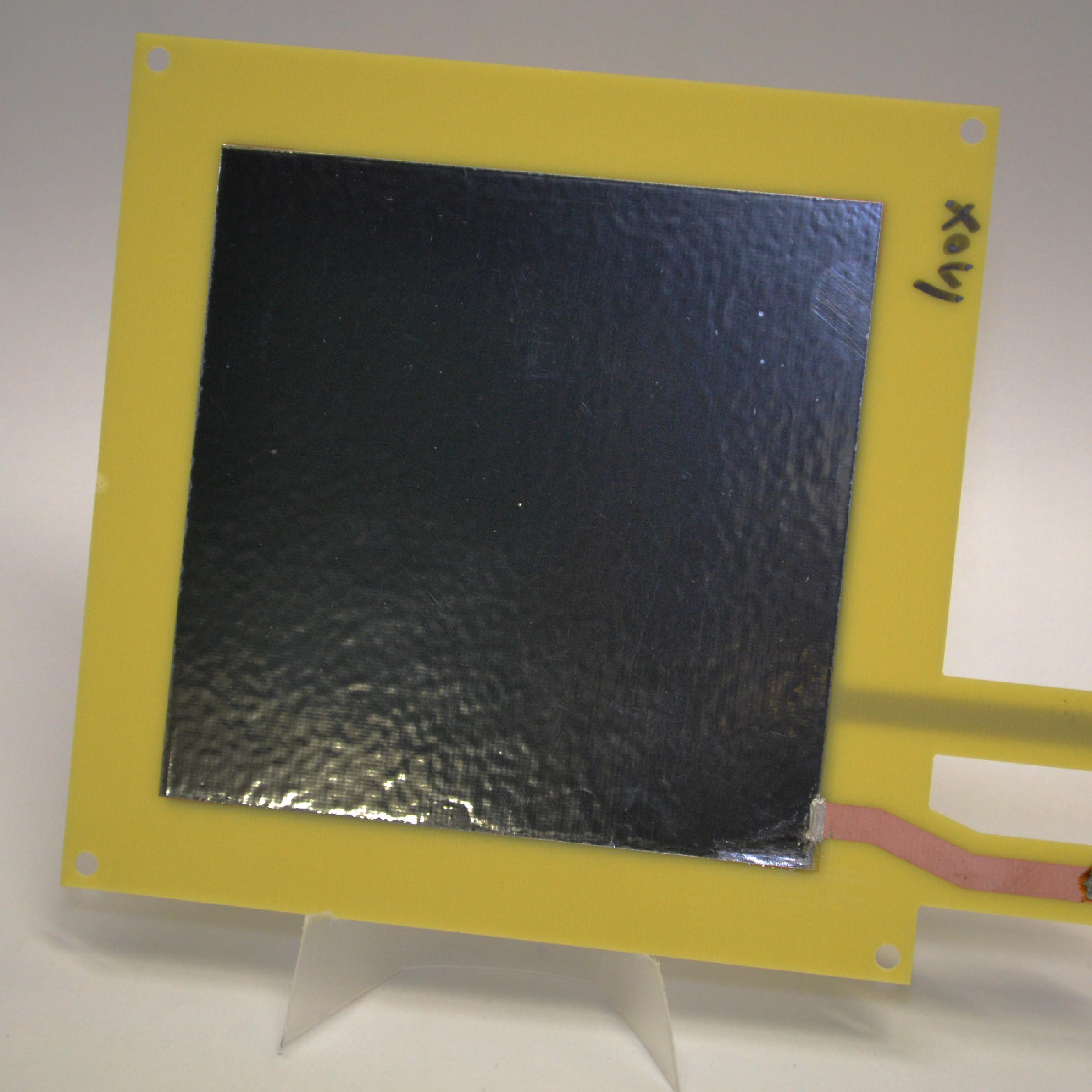}
         \caption{Stainless steel}
         \label{fig:inox_pic}
     \end{subfigure}
     \hfill
     \begin{subfigure}[b]{0.32\linewidth}
         \centering
         \includegraphics[width=\textwidth]{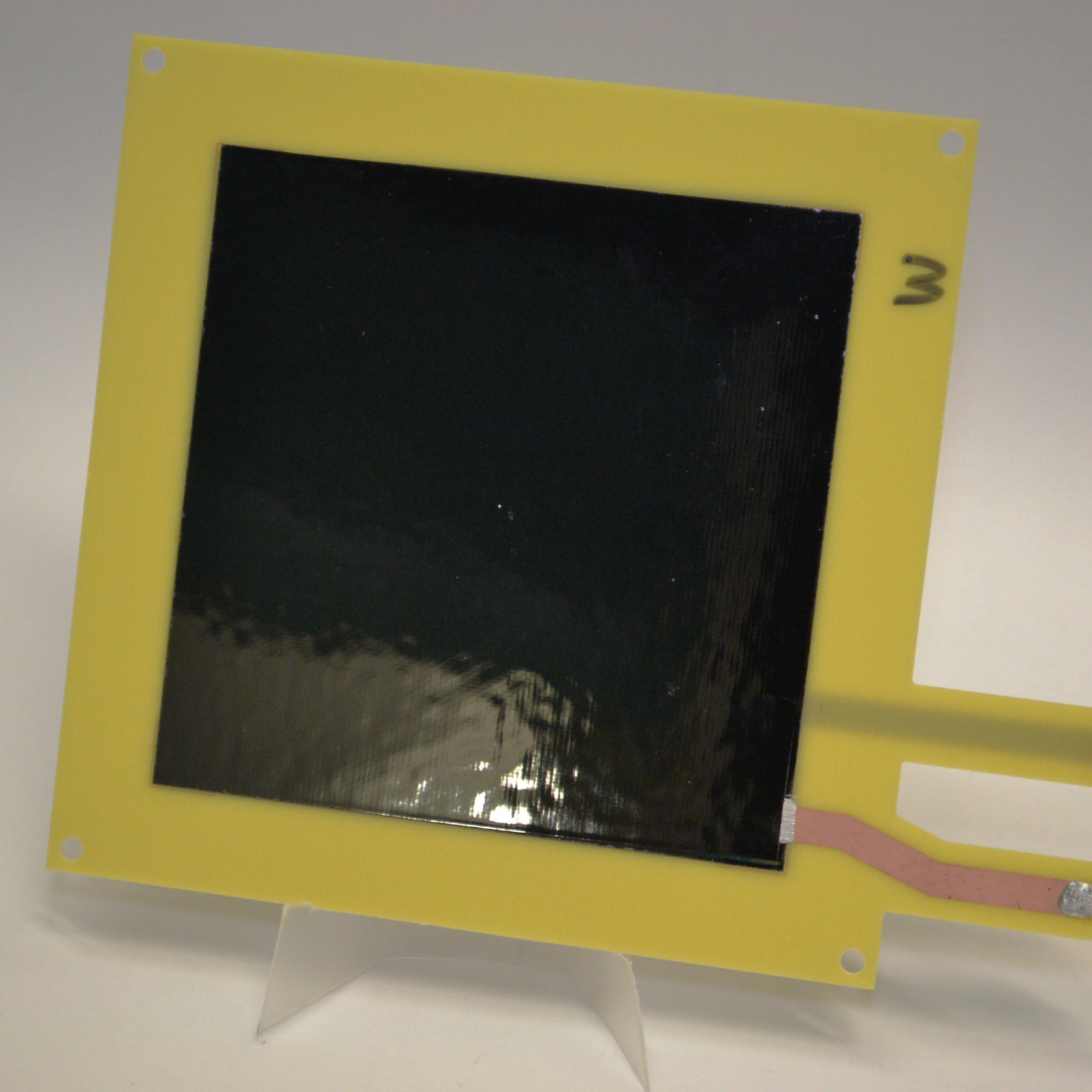}
         \caption{Tungsten}
         \label{fig:w_pic}
     \end{subfigure}
     \hfill
     \begin{subfigure}[b]{0.32\linewidth}
         \centering
         \includegraphics[width=\textwidth]{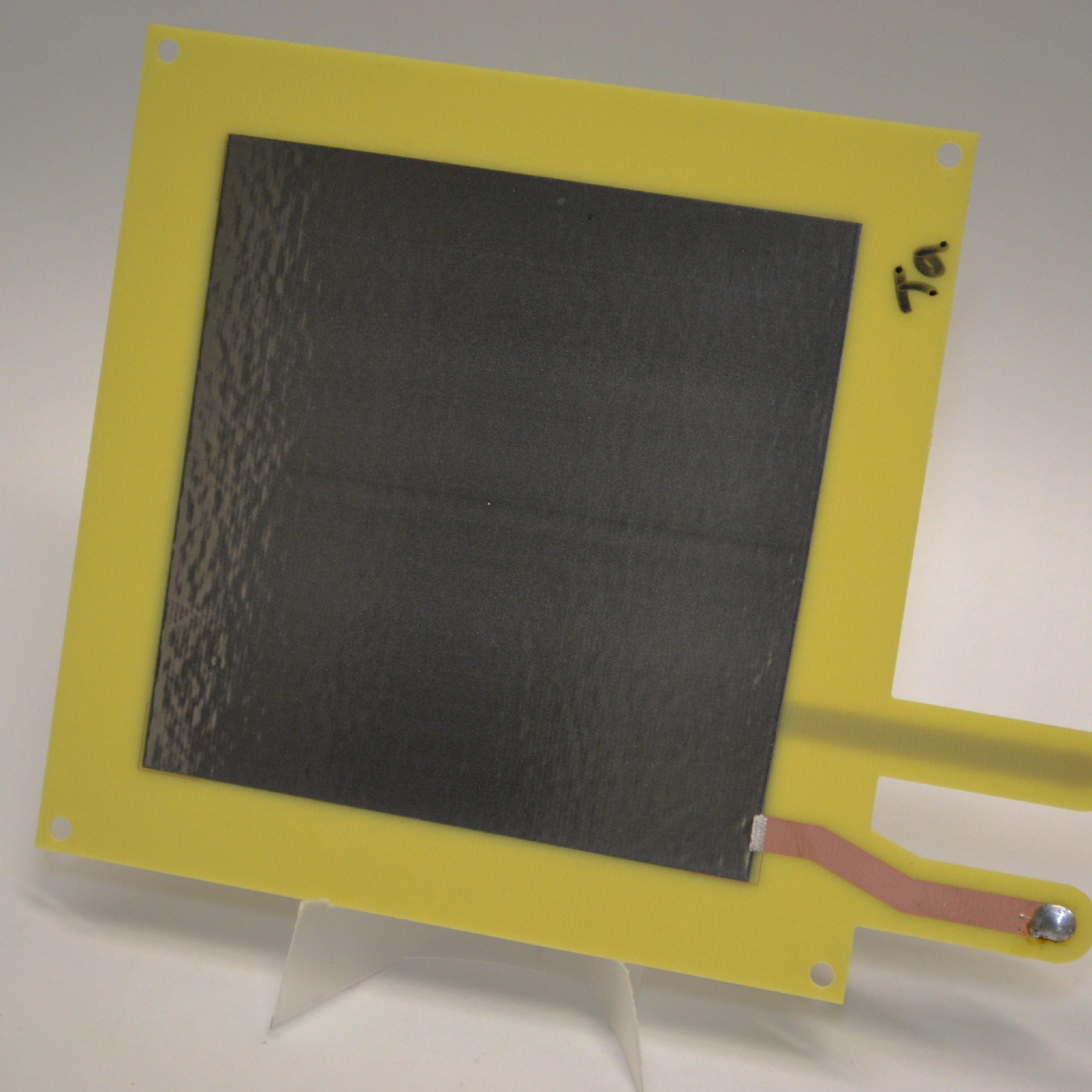}
         \caption{Tantalum}
         \label{fig:ta_pic}
     \end{subfigure}
        \caption{The \sh THGEMs used in the presented studies.}
        \label{fig:THGEM_pics}
\end{figure}

In a standard GEM, both sides of the foil are clad with copper with the resulting conducting layers acting as electrodes. In our studies, we investigate the effects of the choice of material used for this layer. For our measurements, we tested several \sh THGEM and multi-hole GEM foils produced in the CERN PCB workshop which incorporate other metals in place of copper. The tested materials include aluminium (Al), molybdenum (Mo), stainless steel (Inox), tantalum (Ta) and tungsten (W) in addition to copper (Cu). Pictures of the produced \sh THGEMs are shown in \figref{fig:THGEM_pics}.

The produced \sh THGEM foils incorporate a \SI{800}{\micro m} thick FR4 insulating layer sandwiched by \SI{25}{\micro m} thick metal conducting layers on both sides. The metal films are cut from larger sheets to the required 10$\times$10\,cm$^2$ active area. A solid film of cast epoxy is used to adhere the metal films to the base FR4. A thermal hydraulic press is used to cure the epoxy and ensure uniformity. As an exception, the tungsten THGEM incorporates a \SI{50}{\micro m} thick metal layer due to the lack of thinner tungsten sheets. Moreover, the copper THGEM is produced using a standard PCB material which is supplied with a \SI{35}{\micro\meter} copper cladding. After gluing of conductive layers, the THGEM boards are then drilled in the center to create the amplification hole with a diameter of \SI{400}{\micro\meter}. The hole has a cylindrical shape without any rim. The structures are subjected to the standard desmearing process for THGEM cleaning after drilling, including ethylene glycol sweller, permanganate etch and hydrogen peroxide with sulphuric acid finisher. 

The choice of the employed metals was done in a way to cover a large range of thermal and mechanical material properties. The relevant material properties such as conductivity, melting temperature and work function are summarized in Table \ref{tab:material_table}.

\begin{table*}[h!]\footnotesize
\caption{Properties of the coating materials employed in this study (from \cite{toolbox2020materials,derry2015workfunction}).}
\begin{center}
\begin{threeparttable}
\begin{tabular}{ c c c c c c c }
\toprule
Material & Conductivity  & Work function & Melting point  & Boiling point & Thermal conductivity  & Density \\[0.5ex]
        & (\SI{E6}{\siemens\per\meter}) & (\SI{}{\eV}) & (\SI{}{\celsius}) & (\SI{}{\celsius}) & (\SI{}{\watt\per\meter\per\kelvin}) & (\SI{}{\gram\per\centi\meter\cubed}) \\
\midrule
         Al & 36.9 & 4.08 & 660 & 2470 & 237 & 2.702 \\ [0.5ex]
         Cu & 58.7 & 4.7 & 1083 & 2575 & 386 & 8.96 \\[0.5ex]
         Inox & 1.37 & 4.4 & 1510 & 2750\tnote{*} & 16.3 & 7.85 \\[0.5ex]
         Mo & 18.7 & 4.5 & 2623 & 4651 & 138 & 10.22 \\[0.5ex]
         Ta & 7.6 & 4.22 & 3017 & 5365 & 57.5 & 16.65 \\[0.5ex]
         W & 8.9 & 4.5 & 3422 & 5550 & 174 & 19.35 \\
 \bottomrule
\end{tabular}
\begin{tablenotes}
\item[*] value for iron.
\end{tablenotes}
\end{threeparttable}
\end{center}
\label{tab:material_table}
\end{table*}

In addition to the introduced \sh THGEMs, also standard multi-hole GEMs with copper and aluminium cladding are tested in this study. For reliable operation GEMs require intricate patterning of very small features during production. This was not possible to recreate with the mentioned harder metals (Mo, Inox, Ta, W). Which is why only copper and aluminium standard geometry GEM foils were produced.

The two GEM foils incorporate a \SI{50}{\micro m} Apical insulating layer clad with \SI{5}{\micro m} thick copper or aluminium. For the production of the copper GEM, a thin (\SI{0.1}{\micro m}) layer of chromium is used to adhere the copper to the Apical base. Using photo-lithography, a hexagonal hole pattern with a pitch of \SI{140}{\micro m} is marked on the active area of the foil, which in turn is etched to produce the GEM holes. The created double-conical holes have an inner diameter of \SI{50}{\micro m} and an outer diameter of \SI{70}{\micro m}. Similarly, for the production of aluminium GEM, the same process as for the standard copper-cladded foil is used to create the hole structure in the Apical layer. Afterwards, however, the copper and chromium are completely removed and a \SI{10}{\micro m} layer of aluminium is applied by Physical Vapour Deposition. An acid etch is then used to remove the aluminium from the holes selectively resulting in approximately \SI{5}{\micro m} aluminium electrodes.

\section{Experimental setup and measurement methods}
\subsection{Detector}
\label{sec:methods:setup}

\begin{figure}[ht]
    \centering
    \includegraphics[width=\linewidth]{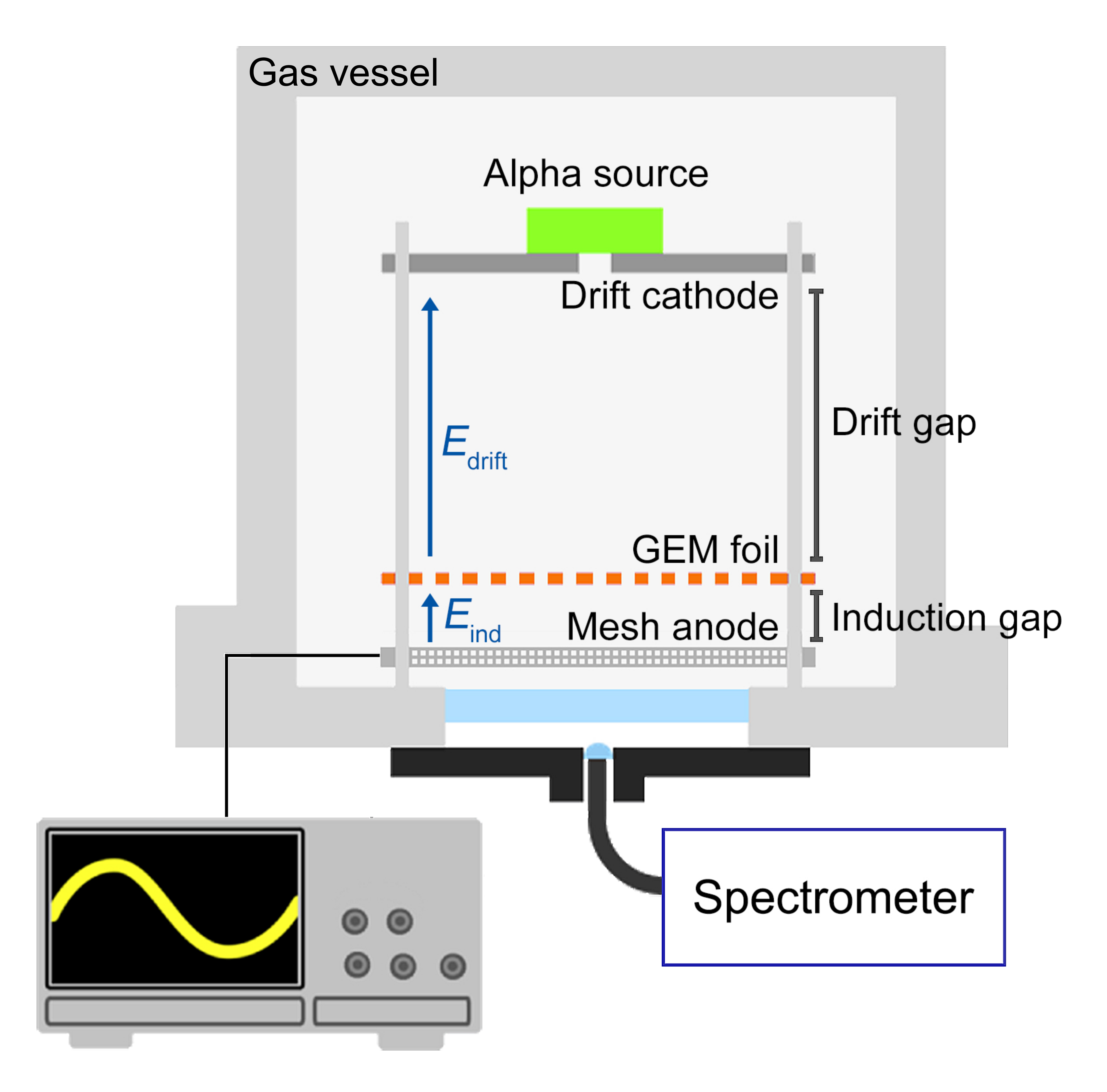}
    \caption{Schematic picture of the experimental setup.}
    \label{fig:spectro_setup_sketch}
\end{figure}

A dedicated detector chamber was built for these studies to facilitate the spectroscopy and secondary discharge stability measurements with the produced GEMs and THGEMs. The setup is schematically shown in \figref{fig:spectro_setup_sketch}. The detector is operated using an independent multi-channel power supply unit and can be flushed with various gas mixtures. Ambient pressure, temperature, and humidity and oxygen levels inside the chamber are monitored during all measurements. For all the conducted discharge and spectroscopy measurements an oxygen contamination below 25\,ppm and a water contamination below 250\,ppmV are maintained.

Inside the chamber, a mixed \alphasource alpha source \cite{alpha} is placed on the top of the cathode electrode facing the drift volume to induce a constant rate of discharges needed for the measurements. The measured alpha rate from the source over the \SI{10x10}{\centi\metre} active area of the used GEMs and THGEMs is \SI[separate-uncertainty = true]{288.1(1)}{Hz}. The drift gap (distance between the top side of the GEM foil and the cathode electrode) is kept at a constant $27$ mm throughout all measurements. The drift electrode facing the top side of the GEM/THGEM foil is connected directly to the HV system. A \SI{5}{M\ohm} protection resistor is used for the top electrode on the GEM foil. No protection resistor is used on the bottom side. The readout anode electrode is facing the bottom side of the GEM foil with a 2 mm induction gap. For spectroscopy measurements, the anode electrode is chosen as a thin mesh with optical transparency of \SI{\sim50}{\%} to maintain a line of sight to the discharging GEM holes. The detector chamber incorporates a BOROFLOAT® window \cite{borofloat} on the bottom side, through which all the discharge light studies are conducted.

\subsection{Emission light spectroscopy}
\label{sec:methods:spectroscopy}
\subsubsection{Spectroscopy measurements}
For the spectroscopy studies, a UV-VIS-NIR Ocean Optics QE65000 \cite{OceanOptics} spectrometer was used with a \SI{600}{\micro\metre} diameter optical fiber connection attached to a collimating lens facing the bottom side of the GEM foil to gather the emitted light during the discharge. The lens is mounted on a separate light-proof housing attached to the detector chamber. The spectrometer is operated with a \SI{10}{\second} exposure duration during which multiple discharges are recorded within a single measurement. The relatively long exposure time is chosen to compensate for the low light emission intensity of single discharges. The number of discharges that occur during the exposure period are counted and the resulting spectra are normalized accordingly. The spectrometer is calibrated using a reference deuterium lamp with a known emission spectrum to correct the intensity distribution over the recorded wavelength range. Furthermore, a wavelength calibration is also conducted by comparing the position of the observed and identified emission lines from the GEM discharges to values taken from spectroscopy databases \cite{NIST}. Prior to each measurement session, spectra are taken using the same settings but with no discharges to obtain the noise levels inherent in the spectrometer pixels.\\[1ex]

\subsubsection{Light spectra analysis}

\begin{figure}[h]
    \centering
    \includegraphics[width=\linewidth]{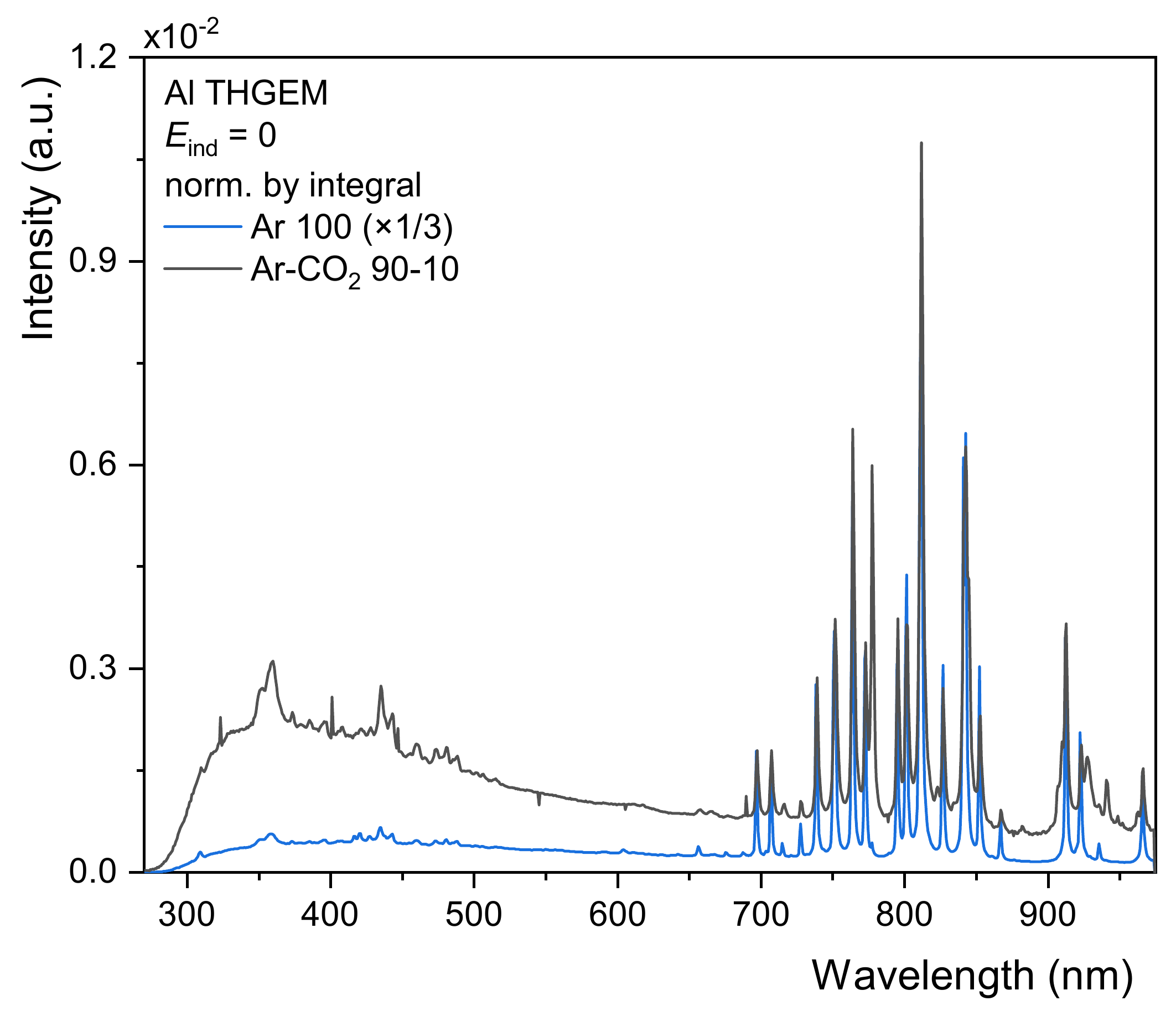}
    \caption{Emission line spectra of Aluminium THGEM discharges measured in \ArCOtwo mixture and pure argon. The blue spectrum is scaled down by a factor of 1/3. See text for details.}
    \label{fig:spectro:quencher}
\end{figure}

An example of a spectrum acquired with the Al THGEM is shown in \figref{fig:spectro:quencher}. Two distinctive features can be associated with the spectrum taken with a mixture of a noble gas (here argon) and a quencher (here CO$_2$). These are the emission lines clearly visible on top of a continuous background. The latter exhibits a characteristic bump in the short wavelength part of the spectrum and can be associated with the molecular emission and absorption processes of the quencher and contamination molecules in the gas mixture \cite{plyler1948co2bands}. The spectrum obtained with the same structure in pure argon shows a significant reduction of the continuous background, especially in the bump region. The remaining background can still be associated with the contaminants present in the gas system such as residual traces of water or oxygen.

In the following studies, we solely focus on the identification of the characteristic emission lines, therefore the background of the spectra can be ignored. For this reason, all spectra shown in \secref{sec:results:gem} are the subject of a background subtraction procedure, carried out to accurately compare the observed emission peaks in different measurements. 

\begin{figure}[ht]
    \centering
    \includegraphics[width=\linewidth]{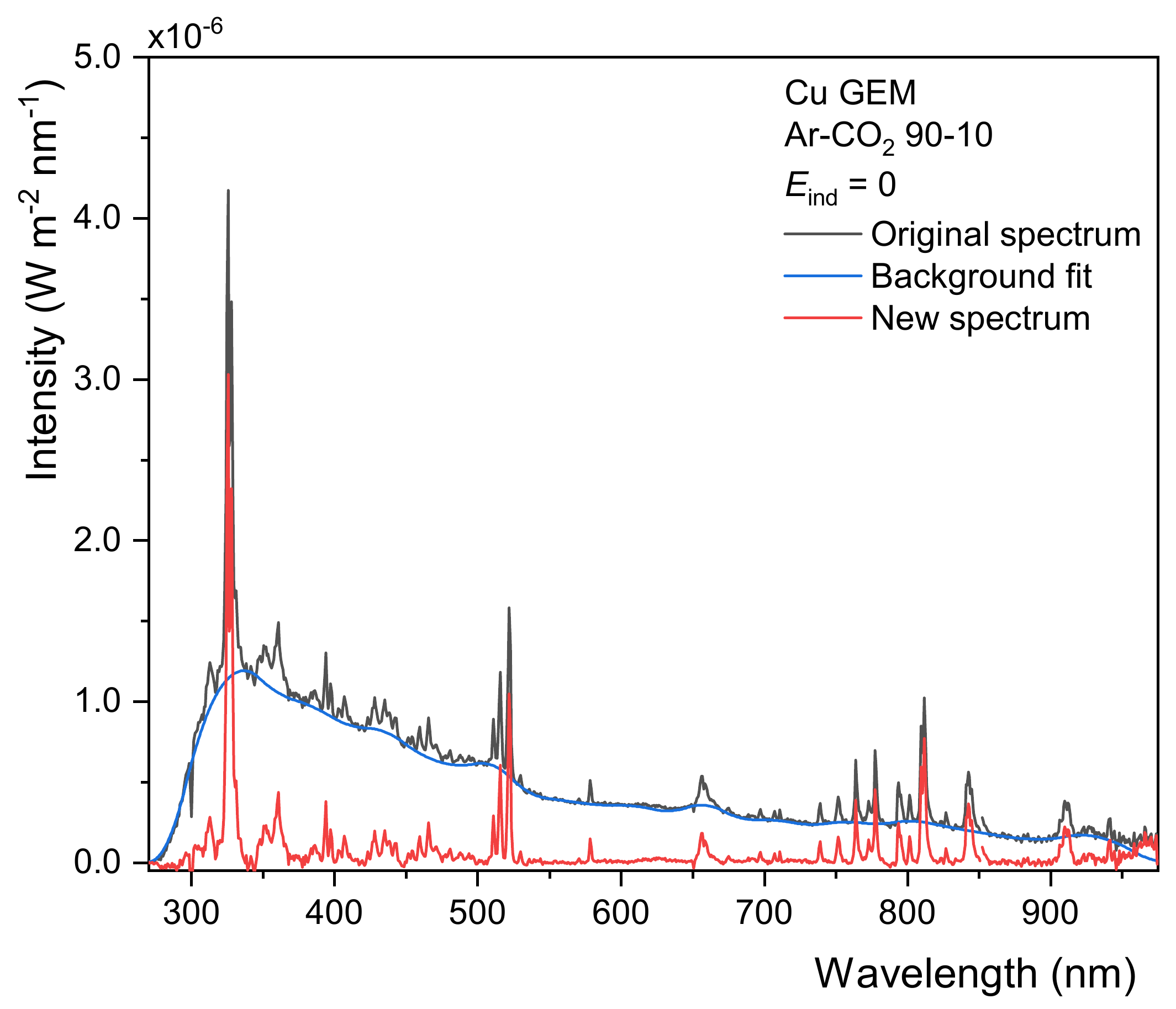}
    \caption{Example of the background removal procedure for a copper GEM discharge spectrum. The continuous background of the original spectrum (black) is fitted (blue) and subtracted. The red plot shows the resulting spectrum after background subtraction. See text for details.}
    \label{fig:spectro_bg_removal}
\end{figure}

An example of the background subtraction is shown in \figref{fig:spectro_bg_removal}. The background distribution is fitted using the algorithm and the tool described and provided in \cite{galloway2009background}. The method employs wavelet transforms to fit the emission peaks in the spectra. In our case the Biorthogonal 3.7 function is selected for this step, as it yielded the best results for our spectra. In the iterative procedure, the algorithm subtracts the fitted peaks above the background level. After ten iterations, the resulting fit distribution converges to the real background distribution. This fit is later subtracted from the original data leading to the background-less spectra. The remaining peaks in these spectra are fitted with a pseudo-voigt function to obtain the mean wavelength value. This is then compared with identified emission lines in spectral databases \cite{NIST} to identify the observed emission lines in the measured spectra. The peak resolution of \SI{\sim1.5}{\nano\meter} (FWHM) can be extracted from the spectra, which is comparable with the optical resolution of the spectrometer given by the producer (\SIrange{0.14}{7}{\nano\meter}, depending on the slit size)~\cite{OceanOptics}.

The final step is the normalization of the overall light intensity to their integral value. It was observed that the mounting structure used for the collimating lens in the setup is prone to small shifts leading to a variance in the absolute light intensity measured on different days. This correction is done in plots denoted with "norm. by integral".

\subsection{Discharge studies}
\label{sec:methods:discharges}

The stability of the produced GEM and THGEM foils against electrical discharges are also measured in addition to the spectroscopy studies. The overall stability of the detector is given by the measurements of primary and secondary discharge probabilities as a function of the applied field and gain configurations. The former is measured as a function of the gain which is determined by the applied voltage across the foil. However, in the case of the \sh THGEM foils the gain can not properly be defined due to having a \sh on a large surface. For this reason, the primary discharge probability in our study is given as a function of the applied voltage across the foil. It is measured by recording the discharge rate for a certain set voltage scheme in a given period and normalizing it to a previously measured source rate. 

It has been shown \cite{deisting2019secondary} that the formation of secondary discharges depends on the applied induction field between the bottom side of the foil and the anode electrode \cite{deisting2019secondary}. Considering these results, the secondary discharge probability ($P_2$), defined as the ratio of the number of secondary discharges to primary discharges, is measured as a function of the applied induction field. The rate measurement of the discharges is done with an oscilloscope connected to the anode electrode. The waveform of the two different types of GEM discharges induced on the anode electrode can easily be distinguished (as can be seen in \figref{fig:primary_secondary}) and counted.
\section{Results}
\subsection{Emission line spectroscopy}
\subsubsection{GEM foils with Al- and Cu-cladding}
\label{sec:results:gem}
\begin{figure}[ht]
    \centering
    \includegraphics[width=\linewidth]{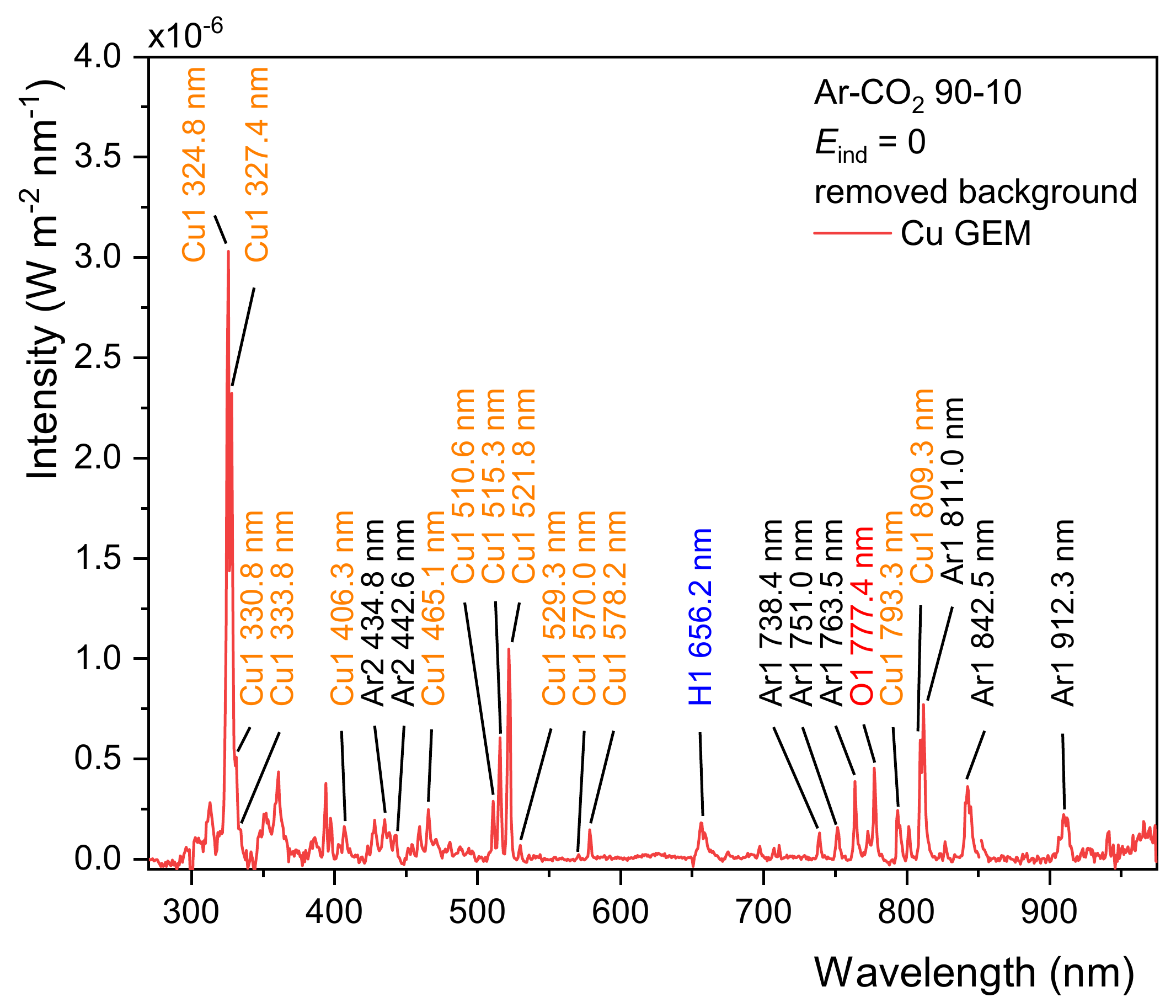}
    \caption{Emission spectra of primary spark discharges in copper GEM detector in \ArCOtwo mixture. Gas components and copper emission lines can be identified, the latter pointing to the vaporization of GEM foil material during a discharge.}
    \label{fig:Cu_GEM_ArCO2}
\end{figure}

The emission spectra from primary discharges of a standard geometry copper GEM  in \ArCOtwo gas mixture are shown in \figref{fig:Cu_GEM_ArCO2}. Copper emission lines can be identified in the shorter wavelength region of the plot, whereas longer wavelengths are dominated by the emission lines from the noble gas. Visible (although weaker) lines of oxygen, carbon and hydrogen can be attributed to the components of the quencher (CO$_2$) and to the contaminants, such as residual amounts of H$_2$O and O$_2$ in the gas (see \secref{sec:methods:setup}). The observation of copper lines points to the vaporization or sputtering of the electrode material during the discharge process. This fact was previously studied indirectly by observing material damages appearing around the GEM holes after many discharges \cite{merlin2016study}. It is important to note that the narrow emission lines visible in the spectrum can only originate from particles being part of the discharge plasma and not contained in the crystal lattice of the bulk foil material. The latter would only be identified in the form of broad bands instead of single lines \cite{griem_1997}. For this reason we conclude that the GEM electrode material is vaporized  during a discharge process.

\begin{figure}[t]
    \centering
    \includegraphics[width=\linewidth]{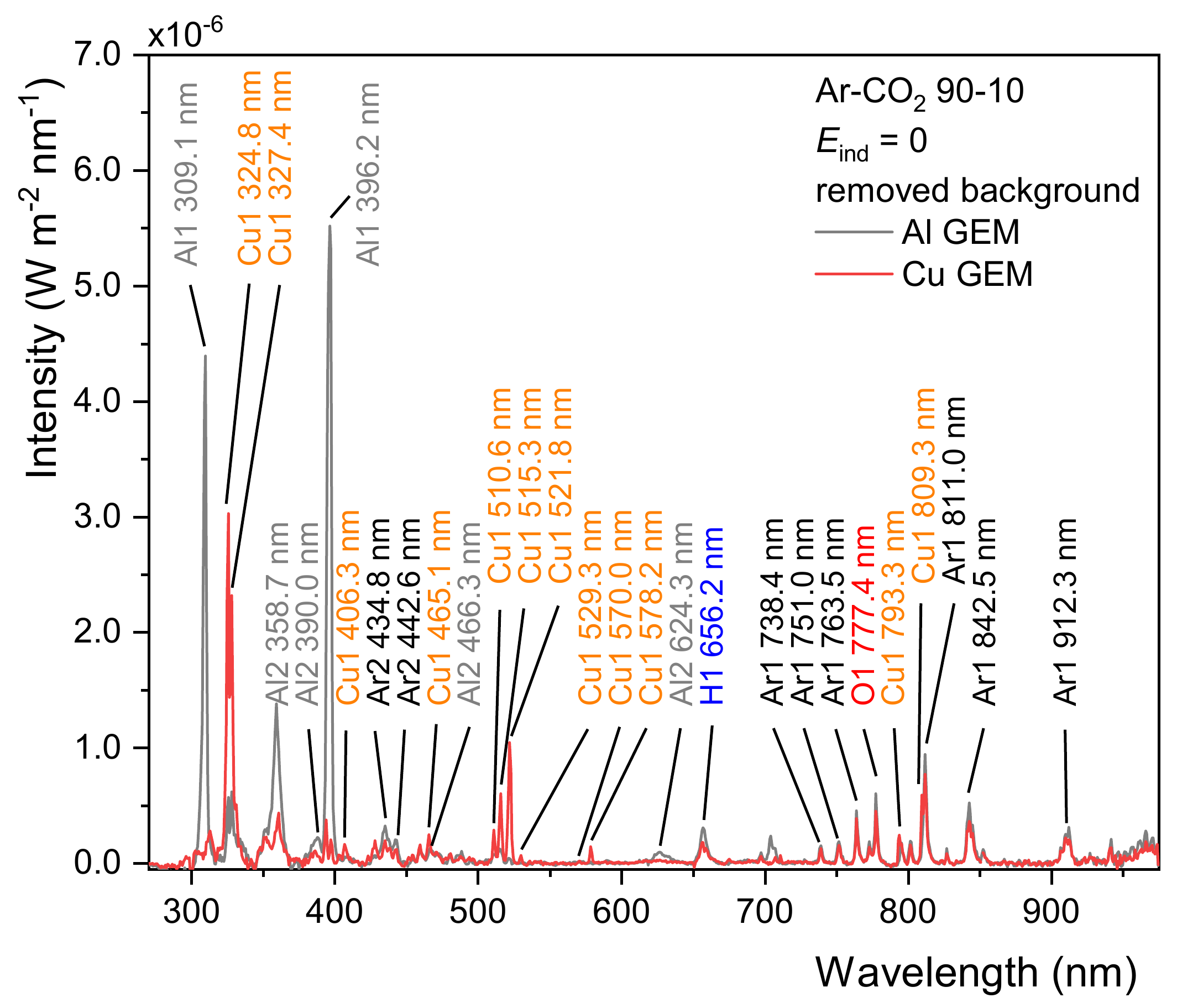}
    \caption{Comparison of emission spectra of primary spark discharges in copper and aluminium standard GEM foils. The emission peaks of the foil materials can be easily distinguished.}
    \label{fig:Cu_vs_Al_spectra}
\end{figure}

\Figref{fig:Cu_vs_Al_spectra} depicts the comparison between the spectra obtained with the copper and aluminium GEM foils. In the case of the aluminium GEM, the peaks attributed previously to copper are not visible. Instead, peaks that match aluminium emission lines appear in the spectrum. On the other hand, peaks attributed to the components of the gas mixture are present in both spectra. 
This gives further confidence that the emission lines are identified correctly and that the used GEM foil cladding material is indeed abundant in discharge plasma. Higher intensity of the aluminium lines, in the measured wavelength range, indicates higher vaporization rate of this element in a discharge process. This observation is supported by the vapor pressure curves of aluminium and copper pointing to higher vapor pressure of the former for a given temperature value~\cite{vapor_pres}. In first approximation, the temperature of an electrode reached in a discharge should be similar for Al- and Cu-cladded GEMs. Both foils have a capacitance $C$ of \SI{\sim5.5}{\nano\farad}, however, the potential difference $\Delta V$ across aluminum GEM is set to 420\,V whereas the corresponding value for the copper GEM is 410\,V. Considering a GEM foil as a capacitor, the energy $E=0.5 C \Delta V^2$ stored in the aluminum structure is $\sim$5\% larger than in the copper one. This, and the lower thermal diffusivity of aluminium may result in slightly higher temperature of the aluminium surface reached in a discharge. This, together with the vapor pressure dependency on temperature, can explain observed relative intensities of the electrode material lines in the measured spectra.

However, without absolute normalisation of the element content in the discharge plasma, it is not possible to extract the exact value of temperature with this method. As the material can vaporize, in principle, at any temperature, one cannot associate the observation of the material lines with any particular temperature value, e.g.~melting point. Moreover, vapor pressure of aluminium at its melting point is almost four orders of magnitude lower than the corresponding value for copper~\cite{vapor_pres}. The boiling point of the material, on the other hand, can provide an upper limit of the possible temperature range. It is therefore useful to study other materials and utilize this method as a temperature gauge of the electrode region around discharges.

\subsubsection{THGEMs with different cladding materials}
\label{sec:res:spect:clad}

\begin{figure*}[ht]
\centering
    \includegraphics[width=0.48\linewidth]{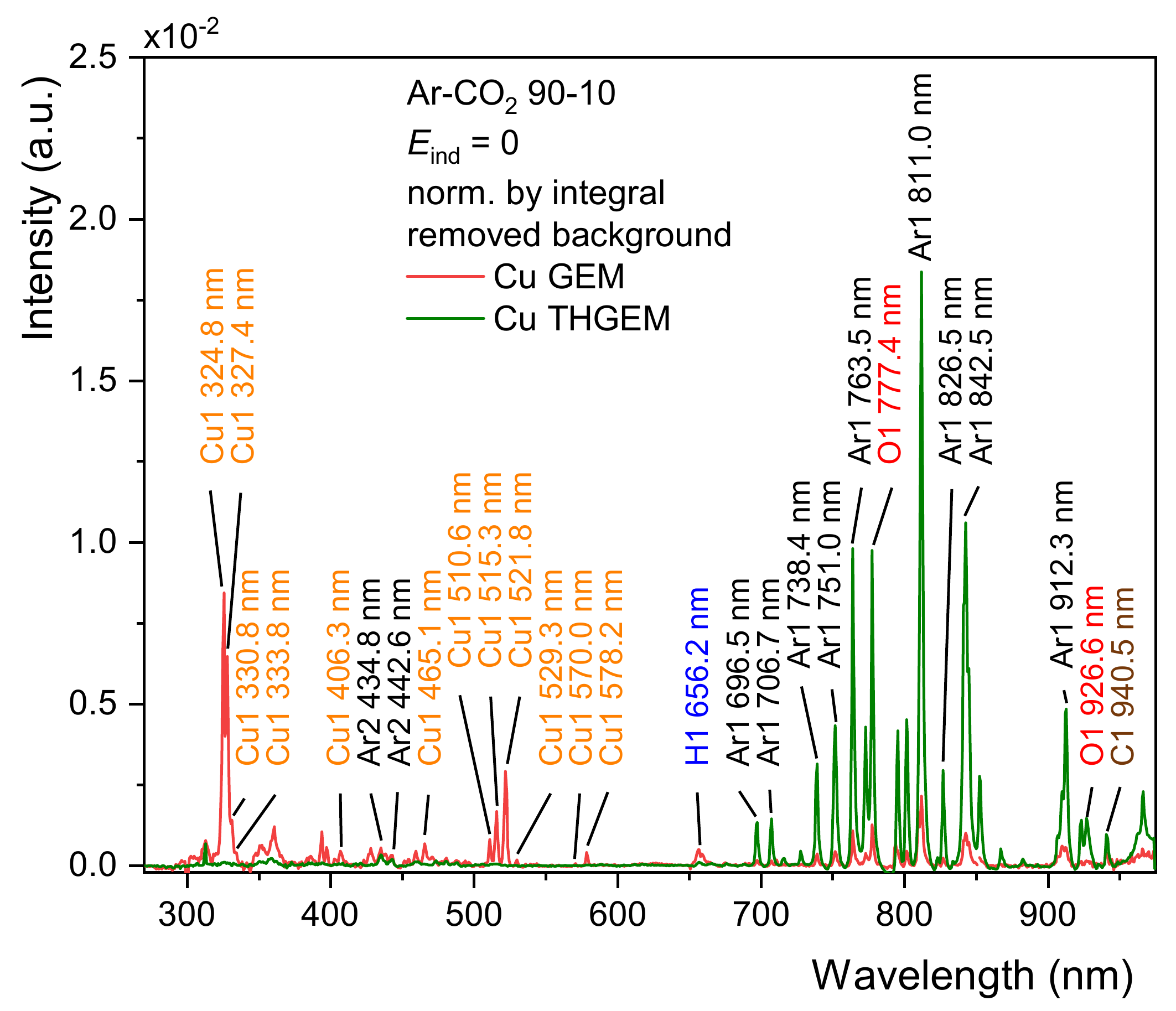}
    \includegraphics[width=0.48\linewidth]{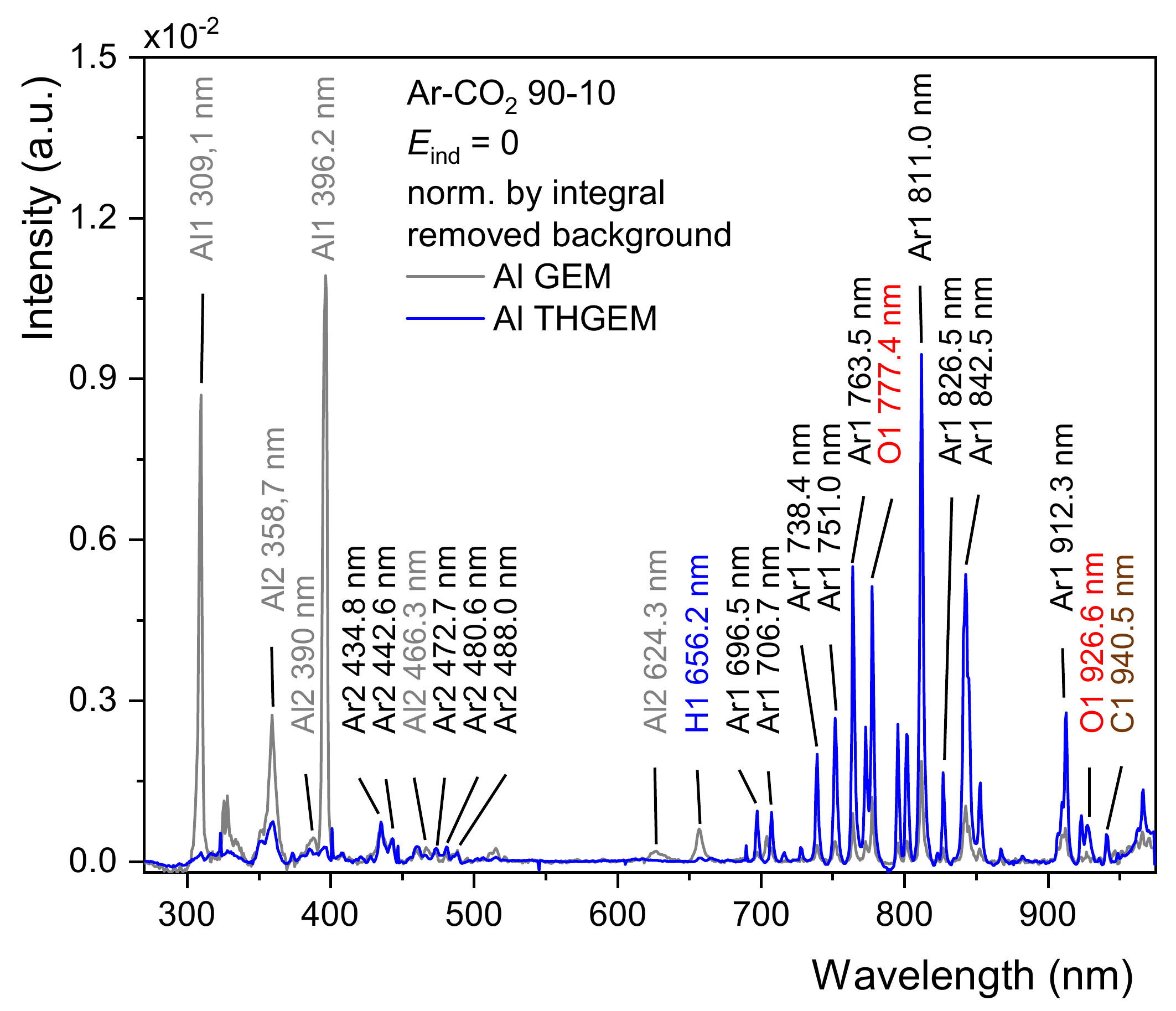}
   
    \caption{Comparison between GEM and \sh THGEM primary spark discharge emission spectra obtained with copper (left) and aluminium (right) cladding.}
\label{fig:GEM_vs_THGEM}
\end{figure*}

Discharge light emission spectra of \sh THGEMs with aluminium and copper electrodes are compared with the corresponding results obtained with GEMs (see \secref{sec:results:gem}) and presented in \figref{fig:GEM_vs_THGEM}. Surprisingly, no material lines are visible in the THGEM spectra apart from a few lines in Al THGEM emerging around 356\,nm and 396\,nm, which may have the same origin as well visible lines in the GEM spectrum. The gas constituent emission lines are, on the other hand, well visible in both, GEM and THGEM, spectra. The absence of foil material lines in the spectra points towards significantly reduced vaporization of the cladding layer. This in turn indicates that the cladding material reaches much lower temperatures. Following the reasoning from \secref{sec:results:gem}, the absence of copper lines and hardly visible signals in aluminium THGEM spectrum indicate that the temperature of THGEM electrodes is most probably lower than $\sim2470^{\circ}$ (boiling point of aluminium). It can be concluded that the thick cladding layer of a THGEM dissipates the heat more efficiently, limiting the maximum temperature reached by the metal around the discharging THGEM. This is even more evident considering the energy stored in a \sh THGEM capacitor. Given the capacitance of \SI{\sim0.55}{\nano\farad}, measured for all THGEMs used in this study, and the potential difference at which the primary discharges are recorded (see \secref{sec:res:primary} and \tabref{tab:primary_onset} for more details), the energy of a \sh THGEM discharge is at least 60\% larger than in a standard GEM foil.  

\begin{figure}[ht]
    \centering
    \includegraphics[width=\linewidth]{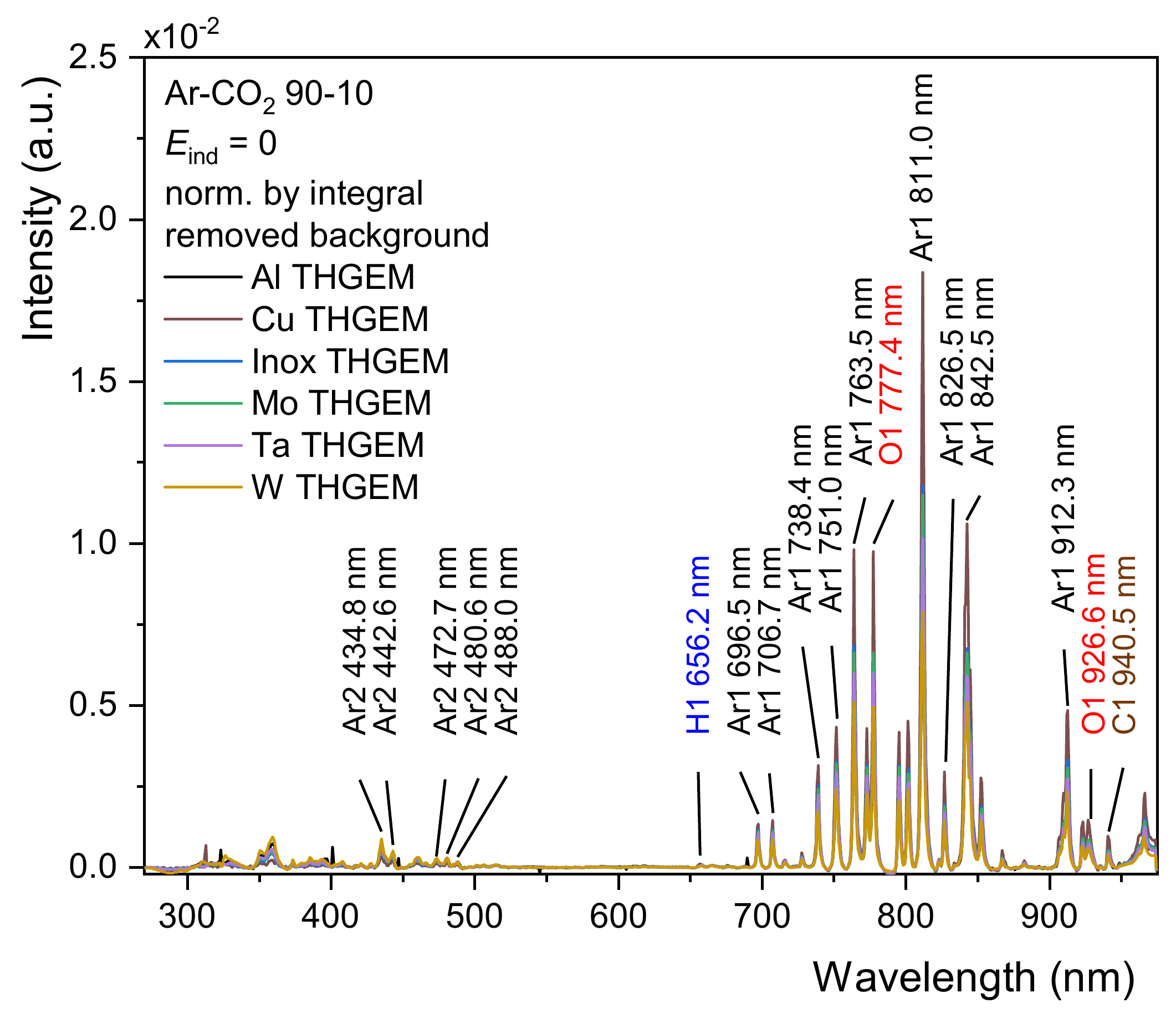}
    \caption{Emission spectra from primary spark discharges measured with \sh THGEM foils with different cladding materials. No emission lines associated with the THGEM materials  can be identified. All spectra share peaks attributed to the gas mixture.}
    \label{fig:Ar100_ExoTHGEMs}
\end{figure}

\break
\Figref{fig:Ar100_ExoTHGEMs} shows the spectra for all THGEMs produced with different materials, indicated in \tabref{tab:material_table}. Similar to the results described in the paragraph above, none of the spectra exhibits emission lines corresponding to the material of the cladding layer. The only lines which can be identified are the emission lines of the employed gas mixture. This points to the significant reduction of material vaporization in any of the studied THGEMs, apart from the possible case of Al THGEM, discussed before, which is compatible with the other materials having even higher boiling temperatures than aluminium. 

\subsubsection{Emission spectra obtained with secondary discharges}

In the last part of the spectroscopy measurements, light from secondary discharges is studied to determine how the two types of discharges differ in terms of their emission spectra. 

\begin{figure}[ht]
    \centering
    \includegraphics[width=\linewidth]{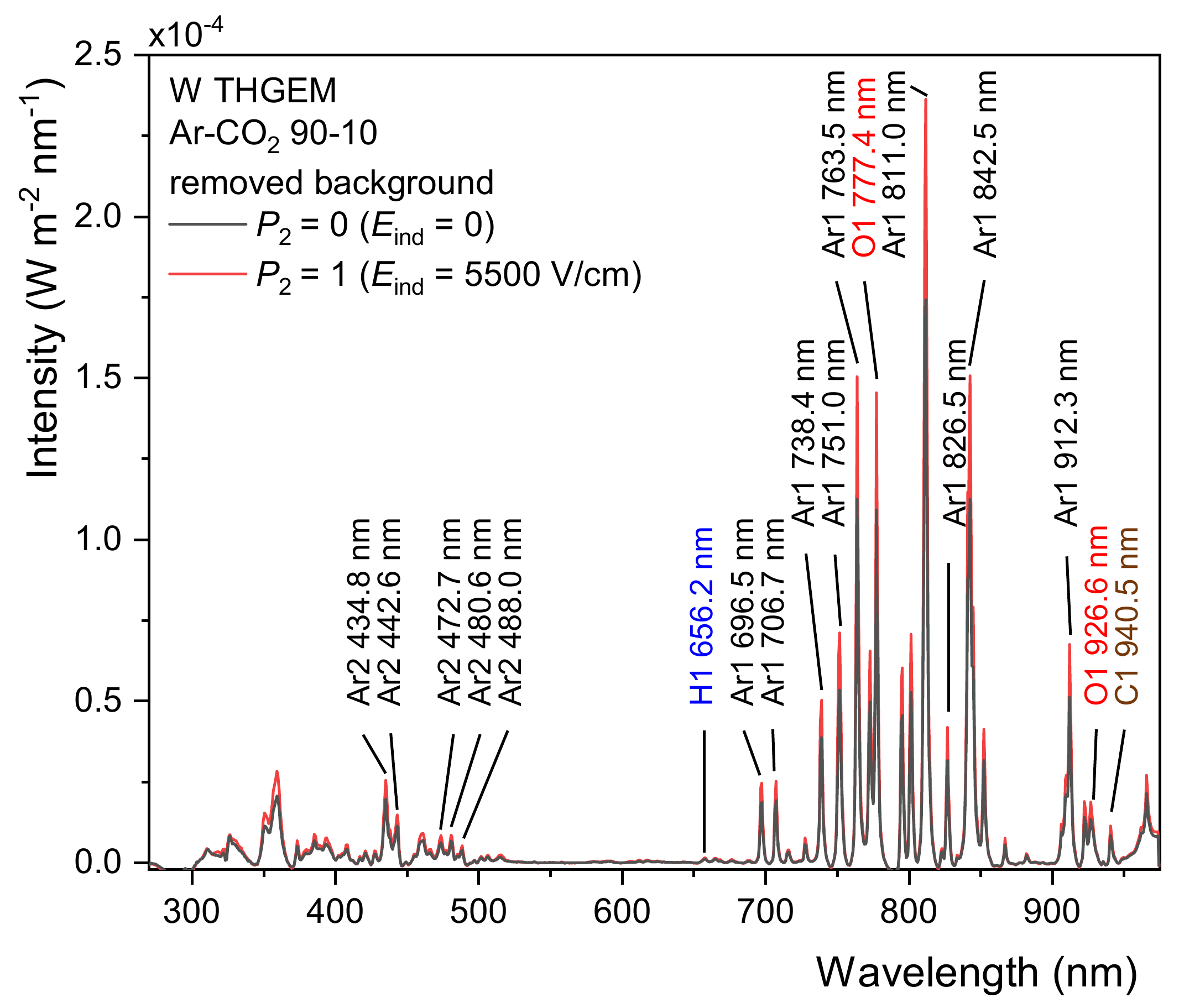}
    \caption{Comparison of emission spectra between events with only a primary spark discharge and events where both primary and secondary discharges are recorded, measured with the tungsten THGEM. Except for the variance in overall intensity, the shape and the peak structure of the spectra are identical.}
    \label{fig:primary_vs_secondary}
\end{figure}

\Figref{fig:primary_vs_secondary} shows a tungsten THGEM spectrum of primary discharge light (black) overlapping with a spectrum of light emission from primary and secondary discharges (red). The latter was taken at the induction field values assuring 100\% probability of the secondary discharge occurrence. As the secondary discharges can only be triggered by the primary sparks (see \secref{sec:intro}) it is not possible to capture the light from secondary discharge exclusively using the long exposure method, described in \secref{sec:methods:spectroscopy}. Still, it is clearly seen, that the light emitted from the secondary discharges doesn't considerably affect the total spectrum shape. No extra lines can be identified and the intensity of all visible lines is slightly increased with respect to the results obtained with primary discharges only, as expected. This indicates there are no significant differences in light emission from primary and secondary discharges. In both cases, these are spark discharges, although their underlying triggering mechanism may still differ, which is discussed in \secref{sec:discussion}.

It needs to be noted, that the secondary discharges can be triggered with all kinds of GEMs and THGEMs studied in this work (see also \secref{THGEM_secondary}), independently from the cladding material. The results obtained with different THGEM electrodes all lead to the same observation. 

\subsection{Discharge stability dependence on cladding material}

The material dependency on the stability of the detectors against primary and secondary discharges is presented in the following. Both primary and secondary discharge probability measurements have been conducted with all kinds of \sh THGEM structures introduced in \secref{sec:GEMs}. It shall be noted, however, that due to the defect developed in the aluminium GEM, comprehensive discharge studies of multi-hole GEM structures have not been performed. 

\subsubsection{Primary discharge probability}
\label{sec:res:primary}

Due to the fact, that the gain evaluation of a \sh structure is not possible with the employed setup, only an onset voltage for primary discharges is measured. The latter is defined as the voltage applied across a THGEM at which the primary discharge rate reaches a value of $\sim$0.2\,Hz, measured with 50\% uncertainty. The results for all THGEM cladding materials are presented in \tabref{tab:primary_onset}. No significant difference is observed between the onset voltages measured for different cladding materials. As the geometry and dimensions of holes in all produced THGEMs are the same (see \secref{sec:GEMs}), the gain of each structure is expected to be similar for a given voltage. Thus, we conclude that there is no fundamental difference between the studied materials on their influence on primary discharge stability. 
This is well in line with the discharge formation mechanism, where the driving factor triggering streamer formation is a total number of charges collected in a single hole \cite{bachmann2002discharge,gasik2017charge}. Hence, with constant primary charge density, the probability of the streamer formation will depend on the hole geometry, not on the material of the electrode. Given the same hole geometry in all foils, no difference in their stability is expected.  

A potential influence of the material is foreseen for restive electrodes, which may create a quenching mechanism by a local field reduction. However, the conductivity values of all electrodes studied in this work (see \tabref{tab:material_table}) are high enough to assume that the resulting resistance introduced in the circuit is negligible in comparison to the 5\,M$\Omega$ protection resistor used in the setup (see \secref{sec:methods:setup}). 

\begin{table}[ht]\footnotesize
\caption{The operational limit for the used THGEM foils before the onset of primary discharges.}
\begin{center}
\begin{tabular}{ c c }
\toprule
\Sh THGEM & Onset voltage of primary \\[0.5ex]
cladding material & discharge formation (V) \\
\midrule
Al & 1690 \\ [0.5ex]
Cu & 1680 \\ [0.5ex]
Mo & 1700 \\ [0.5ex]
Inox & 1685 \\ [0.5ex]
Ta & 1700 \\ [0.5ex]
W & 1683 \\ [0.5ex]
 \bottomrule
\end{tabular}
\end{center}
\label{tab:primary_onset}
\end{table}

\subsubsection{Secondary discharge probability}
\label{THGEM_secondary}

\begin{figure}[t]
    \centering
    \includegraphics[width=\linewidth]{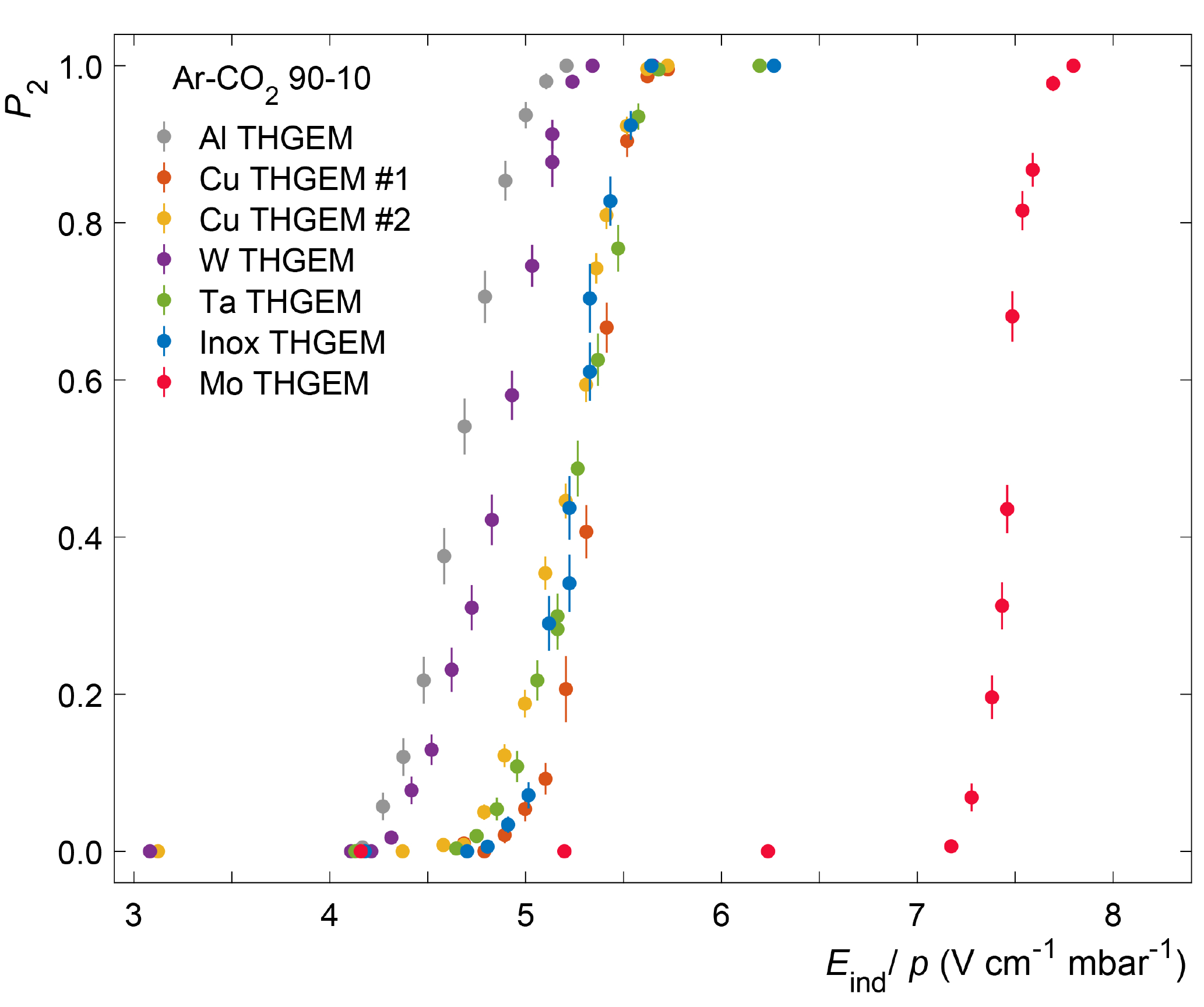}
    \caption{Secondary discharge probability as a function of the induction field measured with \sh THGEMs with different coating materials. Field values are normalised to the ambient pressure of \SI{\sim960}{\milli\bar} measured for each data sample.}
    \label{fig:secondary_discharge}
\end{figure}

\Figref{fig:secondary_discharge} shows the secondary discharge probability $P_2$ measured as a function of the induction field value for all \sh THGEM foils. Secondary discharges could be triggered with all structures. Two sets of results for copper correspond to two different THGEM samples and their variance can be interpreted as a systematic uncertainty of the measurement. 

Clear differences can be observed between the secondary discharge probability curves obtained with different electrode materials, in contrast to the observations made with the primary sparks (see \secref{sec:res:primary}). A group of similar curves can be identified around $E_{\mathrm{ind}}=\SI{5}{\volt\per\centi\meter\per\milli\bar}$ for copper, tantalum and steel THGEMs. 

For tungsten and aluminium, the secondary discharge curves are shifted towards lower fields, pointing to the reduced stability of such structures. It is important to emphasise, that for all the above mentioned materials (aluminium, tungsten, tantalum, steel and copper) the secondary discharges appear, and reach 100\% probability, at the fields where no amplification is expected, i.e.~Townsend coefficient is zero \cite{deisting2019secondary}.

The results can be compared to the measurements with standard copper GEMs in \ArCOtwo reported in \cite{deisting2019secondary}, where the onset and slope of secondary discharge curves were measured at the induction field values between 5 and \SI{6}{\volt\per\centi\meter\per\milli\bar}, for different GEM foils. This shows that the variance in the onset fields between different samples of the same type may be as large as \SI{1}{\kilo\volt\per\centi\meter}, comparable to the spread of secondary discharge curves observed in \figref{fig:secondary_discharge} for the first six THGEM structures. 
On the other hand, it may point to the increased stability of standard copper GEMs (with the onset field of \break \SI{6}{\volt\per\centi\meter\per\milli\bar}) with respect to the aforementioned \sh Cu THGEM. 

Finally, a significant difference can be observed for the molybdenum structure. The discharge curve is shifted towards much higher fields, close to the values, where the Townsend coefficient becomes larger than zero but is still well below the critical field in argon, given by the Paschen's law. It is clear, however, that the stability of the THGEM is substantially improved, when using molybdenum electrodes, even without optimisation of the HV scheme, proposed as a method to mitigate secondary discharges \cite{deisting2019secondary, lautner2019high}. 

\begin{figure}
     \centering
     \begin{subfigure}[b]{0.31\linewidth}
         \centering
         \includegraphics[width=\textwidth]{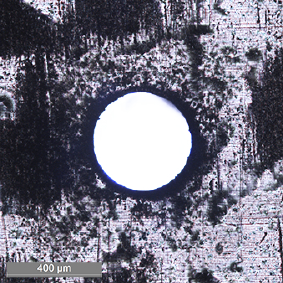}
         \caption{Aluminium}
         \label{fig:alu_pic_after}
     \end{subfigure}
     \begin{subfigure}[b]{0.31\linewidth}
         \centering
         \includegraphics[width=\textwidth]{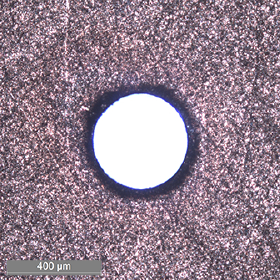}
         \caption{Copper}
         \label{fig:cu_pic_after}
     \end{subfigure}
     \begin{subfigure}[b]{0.31\linewidth}
         \centering
         \includegraphics[width=\textwidth]{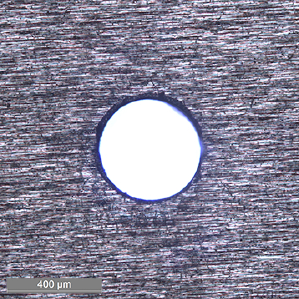}
         \caption{Molybdenum}
         \label{fig:mo_pic_after}
     \end{subfigure}\\
     \begin{subfigure}[b]{0.31\linewidth}
         \centering
         \includegraphics[width=\textwidth]{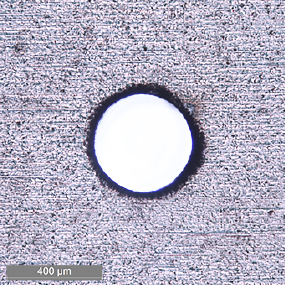}
         \caption{Stainless steel}
         \label{fig:inox_pic_after}
     \end{subfigure}
     \begin{subfigure}[b]{0.31\linewidth}
         \centering
         \includegraphics[width=\textwidth]{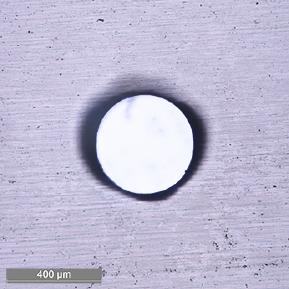}
         \caption{Tungsten}
         \label{fig:w_pic_after}
     \end{subfigure}
     \begin{subfigure}[b]{0.31\linewidth}
         \centering
         \includegraphics[width=\textwidth]{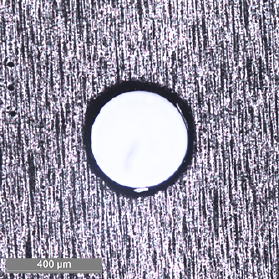}
         \caption{Tantalum}
         \label{fig:ta_pic_after}
     \end{subfigure}
        \caption{Microscope pictures of the THGEM holes (bottom side, facing the induction gap) taken after the measurement campaign.}
        \label{fig:micro_after}
\end{figure}

The holes of all THGEM detectors were studied under an optical microscope to exclude any local defects or differences in geometry, which could point to different discharge performance of the foils. \Figref{fig:micro_after} shows microscope pictures of all six THGEM holes taken after the discharge measurements. No obvious signature can be found, however, for future reference, it might be of interest to investigate the samples using also an electron scanning microscope. 

\subsubsection{Time delay between primary and secondary discharges}
\label{THGEM_delay}

\begin{figure*}[ht]
    \centering
    \includegraphics[width=0.48\linewidth]{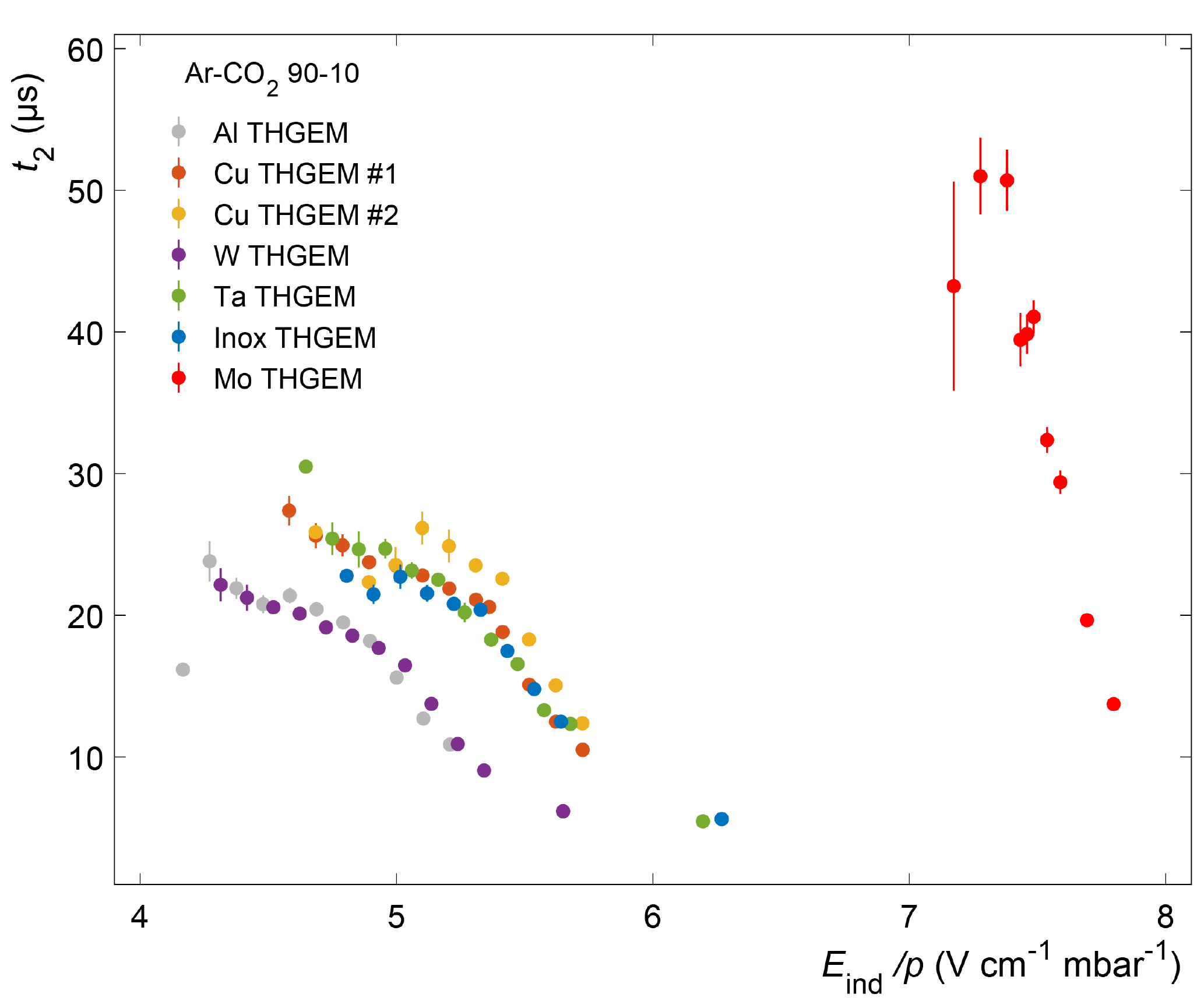}\hspace{2mm}
    \includegraphics[width=0.48\linewidth]{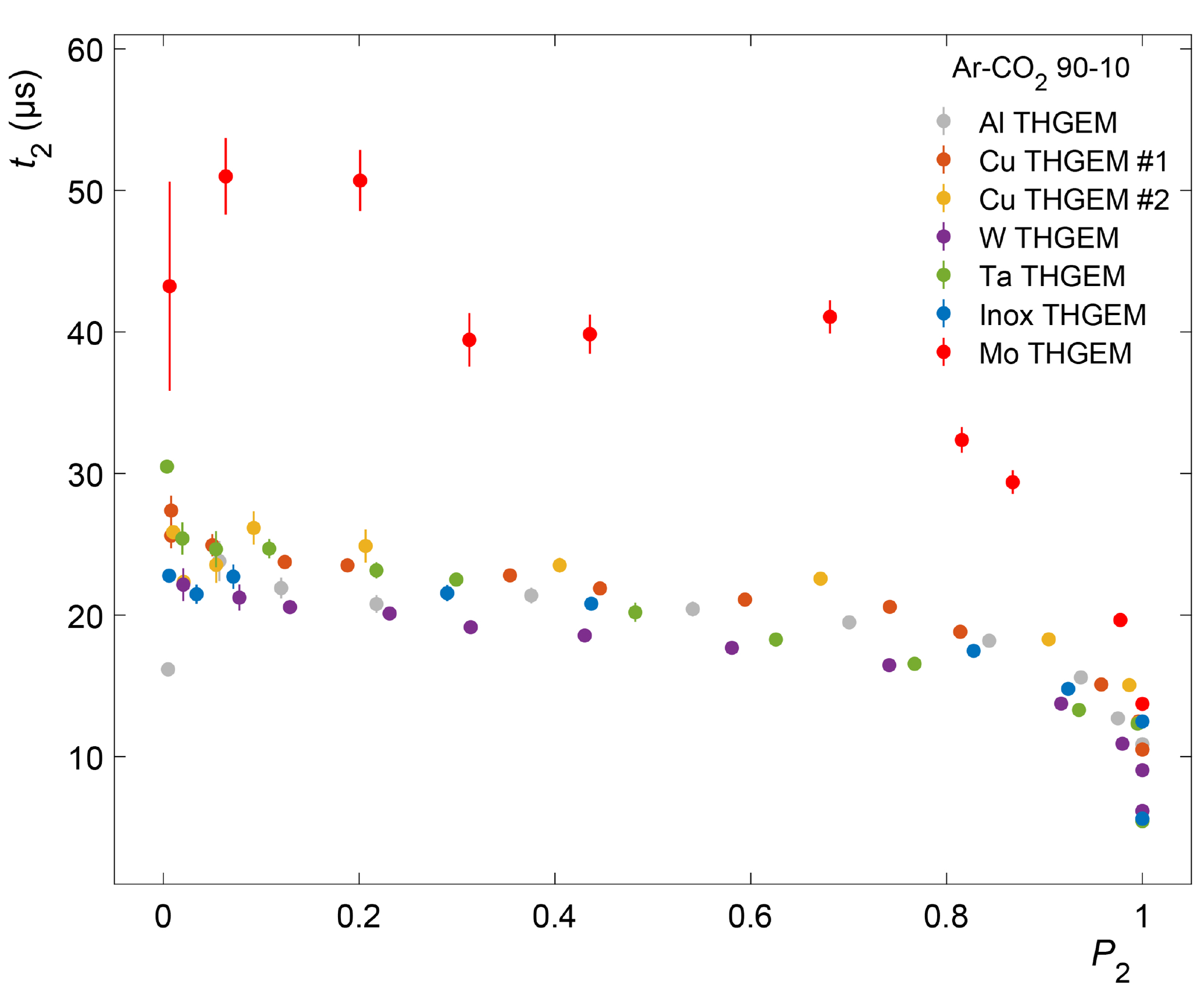}
    \caption{Average time delay between primary and secondary discharges as a function of normalized induction field (left) and secondary discharge probability (right). Error bars indicate the Standard deviation of the mean.}
    \label{fig:secondary_time}
\end{figure*}

An interesting observation can also be made looking at the time delay ($t_2$) between primary and secondary discharges. It is defined, as shown schematically in \figref{fig:primary_secondary}, as the time difference between onsets of both signals measured at their steep, negative slopes. \Figref{fig:secondary_time} shows the average time delays measured for all THGEM types used in this study as a function of the induction field and the secondary discharge probability.

As already reported for Cu GEMs in~\cite{deisting2019secondary}, the time delay of $\mathcal{O}$(\SI{10}{\micro\second}) drops with increasing induction field. The values of $t_2$ are roughly compatible with the drift time of ions in the induction gap. However, their dependency on the $E_{\mathrm{ind}}$ value cannot be explained by the corresponding dependency of ion mobility in \ArCOtwo, reported in~\cite{DEISTING20181}. This suggests that the mechanism responsible for secondary discharge development is more complex than cathode emission caused by the bombardment of positive ions drifting in the induction gap.

The time delays measured for different THGEMs are similar, starting from \SI{\sim25}{\micro\second} at lower discharge fields, reaching values below \SI{10}{\micro\second} at high $E_{\mathrm{ind}}$. The only difference can be observed, again, for molybdenum electrodes. Although $t_2$ for the highest fields, at which secondary discharges are measured, is also \SI{\sim10}{\micro\second}, the time delay at lower fields is significantly longer. The difference is even more pronounced in the right panel of \figref{fig:secondary_time}, where the time delay is plotted as a function of secondary discharge probability. At the intermediate $E_{\mathrm{ind}}$ field values, where $P_2$ does not reach 100\%, the time lag of the secondary discharge is roughly a factor of two higher than for all other THGEMs. This confirms the observation of distinctive properties of the Mo-coated THGEM.
\section{Discussions}
\label{sec:discussion}

The main motivation of the presented work is to further study the mechanism of secondary discharge creation by employing amplification structures with different electrode materials. The results on primary spark development, presented in \secref{sec:res:primary}, are in line with the well established streamer theory of spark discharge \cite{loeb1939fundamental, doi:10.1063/1.1707290, Raether1939, PhysRev.57.722}, where the charge density and the resulting space charge provide sufficient conditions to develop a conductive plasma channel which eventually leads to a breakdown. This process occurs at the electric field values close to (positive streamer) or larger than the breakdown voltage (negative streamer), given by the Paschen's law. The primary discharge develops fast, which cannot be explained by the secondary emission from the cathode, and does not depend on the cathode material. The latter is clearly seen in \secref{sec:res:primary} where no correlation between the onset of primary discharges and the electrode material is observed.

Such a correlation, on the other hand, is clearly seen in the results on secondary discharge stability, discussed in \secref{THGEM_secondary}. This feature does not comply with the streamer theory of a spark discharge. Together with the significant time lag of up to tens of microseconds \cite{deisting2019secondary} it resembles the classic Townsend (slow) mechanism of a gas discharge \cite{TOWNSEND1900}. On the other hand, the resulting discharge event, observed in the gap below the GEM, is clearly a spark discharge, developing rapidly (see the scope signal in \figref{fig:spectro_setup_sketch}), accompanied by the acoustic signal. In addition, as it was shown in previous works \cite{deisting2019secondary, lautner2019high}, it highly depends on the gas composition. This mixture of different discharge signatures points to a mixture or a transition between Townsend (slow) and streamer (fast) discharge mechanism. Such a transition mechanism may appear at the fields below the critical electric field \cite{SF6}. In the case of a GEM-like structure, a secondary discharge process would be triggered by the Townsend avalanche, described with the Townsend first ionization coefficient, initiated by electrons from a primary discharge. Due to the lower than critical electric field in the gap, a streamer development is unlikely. 

However, as shown in~\cite{deisting2019secondary}, the number of electrons stemming from a primary discharge with an energy high enough to ionize gas components (Ar, CO$_2$) may reach a value of $\sim$10$^7$. These electrons drift in the induction gap producing ions, which leads to the space charge accumulation close to the anode (here the readout electrode). The drift of charges in the induction gap was reported and discussed in \cite{deisting2019secondary} and is also observed in this study. Those ions which, after certain drift time comparable with the measured time delay, reach the cathode may cause secondary electron emission leading to further ionization in a strong space charge field. If a sufficiently high space charge is built up, the electric field distortions may eventually give a rise to a streamer which can develop in the gap towards the bottom GEM electrode. 

In addition, with higher value of the induction field further effects may appear resulting in a non-linear $t_2$ decrease with increasing $E_{\mathrm{ind}}$. The observed decrease is larger than expected from the ion mobility and drift time considerations (see \secref{THGEM_delay}). This is, however, still under deliberation and no full explanation of the effects is available. The observed increased time lag of secondary discharges measured with the Mo THGEM is also puzzling. It is possible that the time necessary for space charge accumulation is longer as the number of secondary electrons extracted from the cathode is lower. This could also explain no secondary discharge development at lower induction fields, as one needs sufficient time and number of ions for a sufficiently large space charge build up. Only at the highest fields, exceeding the amplification field in a given gas mixture\footnote{first Townsend coefficient in \ArCOtwo exceeds zero at the field of \SI{\sim7}{\kilo\volt\per\centi\meter}~\cite{deisting2019secondary}.}, the avalanche process is rapid enough to produce necessary amount of charges giving a rise to a streamer. This hypothesis, however, shall be further investigated.

The transition model discussed above complies fairly well with the observed signatures of secondary discharges. In addition, it may also explain the observed secondary discharge development with the inverted field in the gap below GEM \cite{deisting2019secondary}. In the latter, the space charge would build close to the bottom side of a (TH)GEM foil and the streamer would develop towards the electrode below GEM, which in this case would play a role of the cathode. Studying secondary discharge time lag with a Mo THGEM and inverted fields should shed light on the baffling $t_2$ results obtained with the Mo THGEM and will be investigated further.  

Coming back to the secondary emission, the proposed mechanism employs GEM electrodes heated by the primary discharge, which in turn facilitates the thermionic emission of further electrons upon ion bombardment and infra-red radiation \cite{deisting2019secondary}. In the studies described in this work, we do not observe a clear dependency of the secondary discharge creation on the temperature of the electrodes. On the contrary, we show that in case of THGEM electrodes, the temperatures reached upon a primary discharge are lower than for GEMs (see \secref{sec:res:spect:clad}), whereas the onset field for the secondary discharge development remains the same, if not lower, as shown for Cu-based electrodes. In addition, differences between secondary discharge curves are not reflected in the emission light spectra discussed in \secref{sec:res:spect:clad}. The intensity of the lines obtained with different THGEMs does not follow the same (or the opposite) order nor the identified lines differ between measurement with different materials.

This raises the question of whether the temperature effects are indeed crucial for understanding the mechanism of secondary discharge development. On the other hand, we clearly see the dependence on the electrode material, reassuring that the secondary emission plays an essential role in these considerations. It is, however, surprising, that no correlation could be found with basic material parameters such as work function, which drives the emission of secondary electrons. The latter, however, highly depends on the quality of the metallic surface, including the possible creation of oxide layers, corrosion and operation under high temperatures \cite{secelem} and should be studied in more details. Another hint towards this hypothesis may be the variance of onset fields for secondary discharges measured for different samples of copper GEMs and THGEMs (see \secref{THGEM_secondary}), reaching \SI{1}{\kilo\volt\per\centi\meter} or more \cite{deisting2019secondary}. The observed differences may be explained by the quality of electrode material. The available data is, however, too scarce to conclude whether the copper-based GEM is more stable than copper-based THGEM and whether the hole geometry and thickness of the copper layer have any influence on the secondary discharge development.

It should be noted, that the enhanced stability of the molybdenum THGEM could still be explained in a much simpler way. It may be, that the exceptionally good surface quality of the molybdenum electrode, especially around the rim, helps to avoid sharp metallic tips and edges which lead to high electric field values causing instabilities. The same argument could be, in principle, used for the observed higher stability (higher onset fields) of copper-based GEMs, reported in some measurements \cite{deisting2019secondary}. The quality of the edge of a chemically etched GEM hole shall be better than the drilled one. It is, however, not clear why measurements with different GEM samples yield different results. Following the argument in the previous paragraph, the onset field dependency on surface quality shall be systematically studied in order to make the final assessment of the observed variances. For the same reason, further studies are necessary to confirm the observations with the Molybdenum electrode, described in this work. Although the same results were obtained in several attempts, the studies should be repeated with an independently produced structure, including possibility of producing multi-hole molybdenum THGEM.

If confirmed, molybdenum electrodes may be a valid alternative for standard, copper-based GEMs, for the operation under extreme conditions. Note, the field values of $\sim$7000\,V/cm are far beyond the usual settings at which GEM detectors are operated. However, employing molybdenum-based multi-hole (TH)GEM structures would allow one to apply extreme settings and operate the detector with highly-asymmetric fields around the foils. This could be beneficial for low \ibf operation in, among the others, MPGD-based photon detectors.

\section{Summary}

Several GEM and THGEM foils with various cladding materials have been tested in terms of their discharge light spectra and their overall discharge stability. Emission lines from the GEM electrode material have been observed in the light from primary and secondary discharges, which supports the hypothesis of material evaporation during the breakdown. No electrode material emission lines are detected with the \sh THGEMs. This suggests much less or no material evaporation during the THGEM discharge process. In terms of the overall stability against secondary discharges, THGEMs don't show increased robustness in comparison to the used GEMs. 

Furthermore, a strong variance in secondary discharge stability is observed between THGEMs with different materials. However, this observed hierarchy between the materials doesn't follow any of the material properties expected to be relevant for discharge formation (e.g.\,conductivity, work function, melting temperature). Among the different THGEMs used in the studies, the one cladded with molybdenum is shown to be far superior in terms of stability against secondary discharges. If this observation is confirmed with further samples, including single- and multi-hole structures, molybdenum electrodes may be a valid alternative for standard, copper-based GEMs.

With the presented studies, we have demonstrated that spectroscopy can be a capable tool for studying discharges in MPGDs and that the observed material dependence of secondary discharge stability is likely to play a role in solving the secondary discharge formation puzzle. 

\section*{Acknowledgements}
The studies have been performed in the framework of the RD51 Collaboration. 
The authors wish to thank the following persons for many fruitful discussions, ideas, and help with the measurements and interpretation of their results (in alphabetical order): H.~Appelsh\"{a}user, F.~Brunbauer, C.~Garabatos, E.~Oliveri, V.~Peskov, L.~Ropelewski, F.~Sauli and A.~Ulrich.

This work was supported by the Deutsche Forschungsgemeinschaft - Sachbeihilfe [grant number DFG FA 898/5-1].

\bibliographystyle{./elsarticle-num.bst}
\bibliography{./References.bib}

\end{document}